\documentclass[11pt]{article}
\usepackage{amsmath, amssymb, amsthm, epsfig}
\newcommand{\llq}{\leq}

\newtheorem{theorem}{Theorem}[section]

\newtheorem{proposition}[theorem]{Proposition}
\newtheorem{remark}[theorem]{Remark}
\newtheorem{lemma}[theorem]{Lemma}

\newtheorem{Conjecture}[theorem]{Conjecture}
\newtheorem{definition}[theorem]{Definition}

\title{MOPS: Multivariate Orthogonal Polynomials (symbolically)} 
\author{Ioana Dumitriu, Alan Edelman, and Gene Shuman}

\begin{document}

\maketitle

\abstract{In this paper we present a Maple library (MOPs) for computing 
Jack, Hermite, Laguerre, and Jacobi multivariate polynomials, as well as 
eigenvalue statistics for the Hermite, Laguerre, and Jacobi ensembles of 
Random Matrix theory. We also compute multivariate hypergeometric 
functions, and offer both symbolic and numerical evaluations for all 
these quantities. 

We prove that all algorithms are well-defined, analyze their complexity, 
and illustrate their performance in practice. Finally, we also present a 
few of the possible applications of this library.}

\tableofcontents

\section{Introduction}

\subsection{Motivation} \label{mm}
    
There is no need for us to review the impact that classical orthogonal 
polynomial and special functions theory has had for applications in 
mathematics, science, engineering and computations.  By the middle of the 
last century, handbooks had been compiled that could be found on nearly 
everyone's bookshelf.  In our time, handbooks join forces with mathematical 
software and new applications  making the subject as relevant today as it was 
over a century ago.
     
We believe that the modern day extension of these scalar functions are
the multivariate orthogonal polynomials or MOPs along with their special 
function counterparts.  
%These functions are not as well known, algorithms 
%for their computation have not systematically been studied, and important 
%applications may be being missed.  
%This paper attempts to launch a new 
%effort to correct this.

The multivariate cases are far richer, yet at this time they are
understudied, underapplied, and important applications may be being
missed. Algorithms for their computation have not been studied
systematically, and software suitable for scientific computing hardly
exists.  At this time there are no handbooks, no collection of software,
and no collection of applications, though since April 2004 entries are
being introduced into Eric Weinstein's Mathworld
website\footnote{Mathworld, URL http://mathworld.wolfram.com/}.

Development of such software may thus be seen as a whole area of research 
ripe for study. This paper might be thought of as a first step in this
direction; undoubtedly better software will emerge in time. 

We recall that scalar orthogonal polynomials are defined by a positive
weight function $w(x)$ defined on an interval $I \subset \mathbb{R}$.  We define the inner product
\[
<f,g>_{w}~=~\int_I f(x)g(x) w(x) dx
\]
and the sequence of polynomials $p_0(x),p_1(x),p_2(x),\ldots$,
such that $p_k(x)$ has degree $k$, and such that $<p_i, p_j>_w = 0$ if $i \neq j$. This sequence is the sequence of orthogonal polynomials with respect to the weight function $w(x)$. 

There is a (scalar) complex version of this inner product ($I \subset \mathbb{C}$) where we use $\bar{g}$ instead of $g$; this induces a different set of orthogonal polynomials.

We now define the multivariate version of the inner product, and the corresponding orthogonal polynomials. We take any weight function $w(x)$ defined on a segment $I$, and create an $n$-dimensional weight function which is symmetric in each of its $n$ coordinates, and incorporates a repulsion factor which depends on a ``Boltzmann'' constant $\beta$ (or a temperature factor $\alpha = 2/\beta$) which is not seen in the univariate case:
\begin{eqnarray} \label{general_weight}
W(x_1, \ldots, x_n) = \prod_{1\leq i<j \leq n} |x_i - x_j|^{2/\alpha}~\prod_{i=1}^n w(x_i)~.
\end{eqnarray}

We define multivariate orthogonal polynomials $p_{\kappa}^{\alpha}(x_1, \ldots, x_n)$ with respect to the weight $W(x_1, \ldots, x_n)$. The polynomials are symmetric:
they take the same value for any permutation of the $n$ coordinates $x_i$, and they satisfy
$$\int_{I^n} p_\kappa^{\alpha} (x_1, \ldots, x_n) p_\mu^{\alpha} (x_1, \ldots, x_n)  \prod_{i<j} |x_i-x_j|^\beta \prod_{j=1}^n w(x_i)
dx_1 \ldots dx_n = \delta_{\kappa\mu},$$
where $\kappa$ represents the ``multivariate degrees"
of $p_\kappa^{\alpha}$ (the exponent of the leading term).
  
We begin with our fourth  example: symmetric multivariate Hermite
polynomials.  We take $w(x)=e^{-x^2/2}$, so that the
integral is over all of $\mathbb{R}^n$.  The polynomials are
denoted $H_\kappa^\alpha(x)$.
 Our second and third
examples are $w(x)=x^a e^{-x}$ and $w(x)=x^{a_1} (1-x)^{a_2}$.  These are the Laguerre  $L_\kappa^{\alpha,a}$ and Jacobi $J_\kappa^{\alpha,a_1, a_2}$
polynomials. Special
cases of the Jacobi polynomials are the Chebyshev and Legendre polynomials.

Our first example, the Jack polynomials, generalizes the monomial scalar functions,  $x^k$. These polynomials are orthogonal on the unit circle: $w=1$ and
$I =$ the unit circle in the complex plane. Therefore
$I^n$ may be thought of as an $n$ dimensional torus.  
The orthogonality of the Jack polynomials may be found in formula 
(10.35) in Macdonald's book \cite[page 383]{MacDonald_book}.

Tables \ref{unu}, \ref{doi}, \ref{trei}, and \ref{patru} give the 
coefficients of the Jack, Hermite, Laguerre, and Jacobi in terms of the 
monomial symmetric functions (for the first) and the Jack polynomials (for 
the last three). We take all degrees up to total degree 4 for the Jack 
polynomials, up to total degree 3 for the Hermite polynomials, and up to 
degree 2 for Laguerre and Jacobi; the coefficients can be seen by a simple 
call to the procedures, e.g.\footnote{Note the use of $a$ for $\alpha$ and 
$g$ for $\gamma$ in the calls.},

\footnotesize{ {\mathversion{bold} \begin{eqnarray*}
&>&\mbox{{\bf jack(a}}, [2], '~\!J'); \\
&>&\mbox{{\bf hermite(a}}, [1,1,1], \mbox{{\bf n}}, '~\!C'); \\
&>&\mbox{{\bf laguerre(a}}, [1,1], g, \mbox{{\bf n}}, ~'\!C'); \\
&>&\mbox{{\bf jacobi(a}}, [1], g_1, g_2, \mbox{{\bf n}}, ~'\!C');
\end{eqnarray*}}}

\vspace{-.25cm}

\normalsize
\begin{table}[ht]
\caption{Coefficients of the Jack ``J'' polynomial expressed 
in monomial basis} \label{unu}
\vspace{.5cm}

\parbox[t]{6cm}{\begin{tabular}{c|c} $k=1$ & $m_{[1]}$ \\ \hline 
$J_{[1]}^{\alpha}$ & $1$ \end{tabular}} \hspace{3cm} 
\parbox[t]{6cm}{\begin{tabular}{c|c|c} $k=2$ & $m_{[2]}$ & $m_{[1,1]}$ \\ 
\hline $J_{[2]}^{\alpha}$ & $1+\alpha$ & $2$ \\ \hline 
$J_{[1,1]}^{\alpha}$ & $0$ & $2$ \end{tabular}}

\vspace{.5cm}

\parbox[t]{7cm}{\begin{tabular}{c|c|c|c} $k=3$ & $m_{[3]}$ & $m_{[2,1]}$ & 
$m_{[1,1,1]}$ \\ \hline $J_{[3]}^{\alpha}$ & $(1+\alpha)(2+\alpha)$ & 
$3(1+\alpha)$ & $6$ \\ \hline $J_{[2,1]}^{\alpha}$ & $0$ & $2+\alpha$ & 
$6$ \\ \hline $J_{[1,1,1]}^{\alpha}$ & $0$ & $0$ & $6$ \end{tabular}}

\vspace{.5cm}

\parbox[t]{9cm}{\begin{tabular}{c|c|c|c|c|c} 
$k=4$ & $m_{[4]}$ & $m_{[3,1]}$ & $m_{[2,2]}$ & $m_{[2,1,1]}$ & 
$m_{[1,1,1,1]}$ 
\\ \hline 
$J_{[4]}^{\alpha}$ & $(1+\alpha)(1+2\alpha)(1+3\alpha)$ & 
$4(1+\alpha)(1+2\alpha)$ & $6(1+\alpha)^2$ & $12(1+\alpha)$ & $24$ 
\\ \hline 
$J_{[3,1]}^{\alpha}$ & $0$ & $2(1+\alpha)^2$ & $4(1+\alpha)$ & 
$2(5+3\alpha)$ & $24$ \\ \hline 
$J_{[2,2]}^{\alpha}$ & $0$ & $0$ & $2(2+\alpha)(1+\alpha)$ & 
$4(2+\alpha)$ & $24$ \\ \hline
$J_{[2,1,1]}^{\alpha}$ & $0$ & $0$ & $0$ & $2(3+\alpha)$ & $24$ \\ \hline
$J_{[1,1,1,1]}^{\alpha}$ & $0$ & $0$ & $0$ & $0$ & $24$ \\ \hline 
\end{tabular}}

\end{table}

\vspace{-.25cm}

\begin{table}[ht!]
\caption{Coefficients of the Hermite polynomial expressed in Jack ``C'' polynomial basis} \label{doi}

\vspace{.5cm}

\parbox[t]{5cm}{\begin{tabular}{c|c} $k=1$ & $C_{[1]}^{\alpha}$ \\ \hline $H_{[1]}^{\alpha}$ & $1$  \end{tabular}} \hspace{4cm} \parbox[t]{5cm}{\begin{tabular}{c|c|c|c} $k=2$ & $C_{[2]}^{\alpha}$ & $C_{[1,1]}^{\alpha}$ & $C_{[1]}^{\alpha}$ \\ \hline $H_{[2]}^{\alpha}$ & $1$ & $0$  & $-\frac{n(n+\alpha)}{\alpha}$  \\ \hline $H_{[1,1]}^{\alpha}$ & $0$ & $1$ & $\frac{n(n-1)}{1+\alpha}$ \end{tabular}} 

\vspace{.5cm}

\parbox[t]{7cm}{\begin{tabular}{c|c|c|c|c} $k=3$ & $C_{[3]}^{\alpha}$ & $C_{[2,1]}^{\alpha}$ & $C_{[1,1,1]}^{\alpha}$ & $C_{[1]}^{\alpha}$ \\ \hline $H_{[3]}^{\alpha}$ & $1$ & $0$ & $0$ & $\frac{3(n+\alpha)(n+2\alpha)}{(1+2\alpha)(1+\alpha)}$ \\ \hline $H_{[2,1]}^{\alpha}$ & $0$ & $1$ & $0$ & $-\frac{6(n-1)(n+\alpha)(\alpha-1)}{(1+2\alpha)(2+\alpha)}$ \\ \hline $H_{[1,1,1]}^{\alpha}$ & $0$ & $0$ & $1$ & $\frac{3\alpha(n-1)(n-2)}{(2+\alpha)(1+\alpha)}$  \end{tabular}}
\end{table} 

\vspace{.5cm}

\begin{table}[ht!]
\caption{ Coefficients of the Laguerre polynomial expressed in Jack ``C'' polynomial basis} \label{trei}

\vspace{.5cm}

\parbox[t]{3cm}{\begin{tabular}{c|c|c} $k=1$ & $C_{[1]}^{\alpha}$ & $1=C_{[]}^{\alpha}$  \\ \hline $L_{[1]}^{\alpha, \gamma}$ & $-1$ & $\frac{(\gamma\alpha+n+\alpha-1)n}{\alpha}$ \end{tabular}} 

\vspace{.5cm}

\parbox[t]{5cm}{\begin{tabular}{c|c|c|c|c} $k=2$ & $C_{[2]}^{\alpha}$ & $C_{[1,1]}^{\alpha}$ & $C_{[1]}^{\alpha}$ & $1 = C_{[]}^{\alpha}$ \\ \hline $L_{[2]}^{\alpha, \gamma}$ & $1$ & $0$  & $\frac{2(\gamma\alpha+n+2\alpha-1)(n+\alpha)}{\alpha(1+\alpha)}$ & $ \frac{(\gamma \alpha + n + \alpha - 1) (\gamma \alpha + n + 2 \alpha - 1) n (n + \alpha)}{\alpha^2(1+\alpha)}$  \\ \hline $L_{[1,1]}^{\alpha, \gamma}$ & $0$ & $1$ & $-\frac{2 (\gamma \alpha + n + \alpha - 2) (n - 1)}{ 1 + \alpha} $ & $\frac{(\gamma \alpha + n + \alpha - 1) (\gamma \alpha + n + \alpha - 2) n (n - 1)}{\alpha (1 + \alpha)}$ \end{tabular}} 
%
%\vspace{.5cm}
%\parbox[t]{7cm}{\begin{tabular}{c|c|c|c|c|c|c|c} $k=3$ & $C_{[3]}^{\alpha}$ & $C_{[2,1]}^{\alpha}$ & $C_{[1,1,1]}^{\alpha}$ & $C_{[2]}^{\alpha}$ & $C_{[1,1]}^{\alpha}$ & $C_{[1]}^{\alpha}$ & $1 = C_{[]}^{\alpha}$ \\ \hline $L_{[3]}^{\alpha, \gamma}$ & $-1$ & $0$ & $0$ & $\frac{3(\gamma \alpha + n + 3\alpha - 1) (n + 2\alpha)}{\alpha (1 + 2\alpha)}$ &  $0$ & $-\frac{3(\gamma\alpha + n + 2\alpha - 1) (\gamma \alpha + n + 3\alpha - 1) (n + \alpha) (n + 2 \alpha)}{\alpha^2(1 + 2\alpha) (1 +\alpha)}$ & $\frac{(\gamma \alpha + n + \alpha - 1) (\gamma \alpha + n + 2\alpha - 1) (\gamma \alpha + n + 3\alpha - 1) n (n + \alpha)(n+2\alpha)}{\alpha^3(1 + 2 \alpha) (1 + \alpha)}$ \end{tabular}}
\end{table}

\vspace{.5cm}

\begin{table}[ht!]
\caption{ Coefficients of the Jacobi polynomial expressed in Jack ``C'' polynomial basis} \label{patru}

\vspace{.5cm}

\parbox[t]{3cm}{\begin{tabular}{c|c|c} $k=1$ & $C_{[1]}^{\alpha}$ & $1=C_{[]}^{\alpha}$  \\ \hline $J_{[1]}^{\alpha, g_1, g_2 }$ & $-1$ & $\frac{(g_1\alpha+n+\alpha-1)n}{g_1\alpha+g_2\alpha+2n-2+2\alpha}$ \end{tabular}} 

\vspace{.5cm}

\parbox[t]{5cm}{\begin{tabular}{c|c|c|c|c} $k=2$ & $C_{[2]}^{\alpha}$ & $C_{[1,1]}^{\alpha}$ & $C_{[1]}^{\alpha}$ & $1 = C_{[]}^{\alpha}$ \\ \hline $J_{[2]}^{\alpha, g_1, g_2}$ & $1$ & $0$  & $\frac{2(g_1\alpha+n+2\alpha-1)(n+\alpha)}{(g_1\alpha+g_2 \alpha +2n-2+4\alpha)(1+\alpha)}$ & $ \frac{(g_1 \alpha + n + \alpha - 1) (g_1 \alpha + n + 2 \alpha - 1) n (n + \alpha)}{(g_1 \alpha +g_2 \alpha +2n-2+4\alpha)(g_1\alpha + g_2\alpha + 2 n - 2 + 3 \alpha)\alpha(1+\alpha)}$  \\ \hline $J_{[1,1]}^{\alpha, g_1, g_2}$ & $0$ & $1$ & $-\frac{2\alpha (g_1 \alpha + n + \alpha - 2) (n - 1)}{(g_1\alpha + g_2\alpha + 2 n - 4 + 2\alpha )(1+\alpha)} $ & $\frac{2\alpha(g_1 \alpha + n + \alpha - 1) (g_1 \alpha + n + \alpha - 2) n (n - 1)}{(g_1\alpha + g_2\alpha + 2 n - 4 + 2\alpha)(g_1\alpha + g_2\alpha + 2 n - 3 + 2\alpha) (1 + \alpha)}$ \end{tabular}} 
\end{table}

\subsection{History and connection to Random Matrix Theory}

The Jack polynomials have a very rich history. They represent a family of orthogonal polynomials dependent on a positive parameter $\alpha$, and some of them are more famous than others. There are three values of $\alpha$ which have been studied independently, namely, $\alpha = 2,1,1/2$. The Jack polynomials corresponding to $\alpha = 1$ are better known as the Schur functions; the $\alpha = 2$ Jack polynomials are better known as the zonal polynomials, and the Jack polynomials corresponding to $\alpha = 1/2$ are known as the quaternion zonal polynomials. 

In an attempt to evaluate the integral (\ref{int5}) in connection with the non-central Wishart distribution, James \cite{james60} discovered the zonal polynomials in 1960. 
\begin{eqnarray} \label{int5}
\int_{O(n)} (\mbox{tr}(AHBH^{T}))^k~~(H^{T}dH) = \sum_{\kappa \vdash k} 
c_{\kappa} Z_{\kappa}(A) Z_{\kappa}(B)~.
\end{eqnarray}

Inspired by the work of James \cite{james60} and Hua \cite{hua63}, in his
own attempt to evaluate (\ref{int5}), Jack was lead to define the
polynomials eventually associated with his name \cite{henry_jack}. More
explicitly, Jack orthogonalized the \emph{forgotten symmetric functions} 
\cite[page 22]{MacDonald_book}, using the inner product given 
in Definition \ref{oo}\footnote{The authors would like to thank 
Plamen Koev for an in-detail explanation of this fact.}. 
He studied the resulting one-parameter ($\alpha$) class of polynomials 
(which now bear his name), and for $\alpha = 1$ he proved were 
the Schur functions, while for $\alpha = 2$ he conjectured to be the zonal 
polynomials (and proved his claim in a very special case).

%We illustrate the process employed by Jack to define the Jack 
%polynomials in Diagram ... He 

He consequently generalized the $\alpha$ parameter to any real non-zero
number, and noted that for $\alpha = -1$ he obtained yet another
special class of functions, which he called the ``augmented'' monomial
symmetric functions. Later it was noted that the orthogonalizing inner
product was positive definite only if $\alpha>0$.

During the next decade, the study of Jack polynomials intensified; Macdonald \cite[page 387]{MacDonald_book} points out that in 1974, H.O.Foulkes \cite{foulkes} raised the question of finding combinatorial interpretations for the Jack polynomials. This question was satisfactorily answered in 1997 by Knop and Sahi \cite{knop-sahi}.

In the late '80s, the Jack polynomials were the subject of investigation in Macdonald's book \cite{MacDonald_book} and Stanley's paper \cite{stanley_jacks}; these two authors generalized many of the known properties of the Schur functions and zonal polynomials to Jack polynomials. 

\vspace{.5cm}

As mentioned, an important application of the Jack polynomials came in conjunction with random matrix theory and statistics of the $2/\alpha$-ensembles. Below we mention a few of the researchers who have made significant contributions in this area. 

\vspace{.5cm}

James \cite{james64a} was one of the first to make the connection between the zonal polynomials ($\alpha=2$ Jack polynomials) and the $1$-ensembles, when he calculated statistical averages of zonal polynomials over the $1$-Laguerre ensemble (Wishart central and non-central distributions). 

At about the same time, Constantine and Muirhead provided a generalization of the hypergeometric series, using the zonal polynomials, and studied the multivariate Laguerre polynomials for $\alpha =1$ (for a reference, see \cite{muirhead82a}).

In a survey paper, James defined and described multivariate Laguerre, Hermite and Jacobi polynomials for $\alpha = 1$ \cite{james75}. Chikuse \cite{LAA} studied more extensively the multivariate Hermite polynomials for $\alpha = 1$. 

In the early '90s, Kaneko \cite{kaneko} studied the general $\alpha$ binomial coefficients, and used them in connection with the study of hypergeometric series and multivariate Jacobi polynomials. He also studied Selberg\index{Selberg integral}-type integrals and established the connection with generalized Jacobi polynomials. A few years later, Okounkov and Olshanski \cite{Ok_Osh_shifted} considered shifted Jack polynomials for all $\alpha$, and proved that they were the same as the generalized binomial coefficients.

Kadell \cite{kadell_jacks} was perhaps the first to consider averages of many valued Jack polynomials, with his study of the average of the Jack polynomial of parameter $1/k$ (with $k$ an integer) over the corresponding $2k$-Jacobi ensemble. Later it was noticed that constraining $k$ to be an integer was unnecessary.

Lasalle \cite{lasalle_herm,lasalle_jac,lasalle_lag}, considered all three types of general $\alpha$ multivariate polynomials, and among many other things computed generating functions for them.

The last results that we mention here are those of Forrester and Baker \cite{Forrester_poly}, who studied in detail the multivariate, general $\alpha$ Hermite and Laguerre polynomials, in connection with the $2/\alpha$-Hermite and Laguerre ensembles (some of their work built on Lasalle \cite{lasalle_herm,lasalle_lag}). For a good reference on multivariate generalizations of many of the univariate properties of the Hermite and Laguerre ensembles, see \cite{Forrester_book}.
 
\section{Multivariate Functions: Definitions, Properties, and Algorithms} \label{math}

\subsection{Partitions and Symmetric Polynomials}

\begin{definition}
A partition $\lambda$ is a finite, ordered, non-increasing sequence of positive integers $\lambda_1 \geq \lambda_2 \geq \lambda_3 \geq \ldots \geq \lambda_l$. 
\end{definition}

Throughout this paper, we will refer to $l = l(\lambda)$ as the length of $\lambda$, and to $k = |\lambda| = \sum_{i=1}^l \lambda_i$ as the sum of $\lambda$. 

\begin{remark} Naturally, one can remove the constraint ``finite'' from the definition of the partition, and replace it with ``of finite sum'', since one can always ``pad'' a partition with $0$s at the end; in this context $l$ becomes the index of the smallest non-zero component of the partition $\lambda$.
\end{remark}

We will work with two orderings of the partitions. The first one is the \emph{lexicographic} one, denoted by $\leq$.

\begin{definition}
We say that  $\lambda \leq \kappa$ in lexicographic\index{lexicographical ordering!definition}al ordering if for the largest integer $m$ such that $\lambda_i = \kappa_i$ for all $i < m$, we have $\lambda_m \leq \kappa_m$. If $\lambda_m<\kappa_m$, we say that $\lambda<\kappa$.\end{definition}

\begin{remark} This is a total ordering of the partition\index{partition}s. \end{remark}

The second ordering is the \emph{dominance\index{dominance ordering!definition}} ordering, sometimes also called the \emph{natural} ordering.

\begin{definition}
We say that $\lambda \preceq \kappa$ (or, equivalently, that $\kappa$ ``dominates'' $\lambda$) if, given $m = \max \{length(\kappa), length(\lambda)\}$, 
\begin{eqnarray*}
\sum_{i=1}^j \lambda_i &\leq &\sum_{i=1}^{j} \kappa_i, ~~~~\forall~j < m~,~~~~\mbox{and} \\
\sum_{i=1}^m \lambda_i & = & \sum_{i=1}^m \kappa_i~.
\end{eqnarray*}
If one of the inequalities above is strict, we say that $\lambda \prec \kappa$.
\end{definition}

\begin{remark} Note that we compare two partition\index{partition}s only if they sum to the same integer. Also note that even with this constraint, $\preceq$ is only a partial ordering of the set of partition\index{partition}s of a given number: for example, $[4,1,1]$ and $[3,3]$ are incomparable.
\end{remark} 

The above summarizes what the user should know about partition\index{partition}s in order to use our library.

\begin{definition} A symmetric polynomial of $m$ variables, $x_1, \ldots, x_m$, is a polynomial which is invariant under every permutation of $x_1, \ldots, x_m$.
\end{definition}

\begin{remark} The symmetric polynomials form a vector space over $\mathbb{R}$.
\end{remark}

Over the course of time, combinatorialists have defined a variety of \emph{homogeneous} bases for this vector space; each such basis is indexed by partition\index{partition}s (which correspond to the terms of highest order in lexicographic\index{lexicographical ordering}al ordering of the polynomial). By homogeneity we mean that all terms of a polynomial in the basis have the same total degree (but this degree varies from polynomial to polynomial).

Some of these homogeneous bases are displayed in the table below: 

\vspace{.25cm}

\begin{center}
\begin{tabular}{|l|l|l|} \hline
Name & Definition for $l=1$ & Definition for $l>1$ \\ \hline
power-sum\index{symmetric functions!power-sum} functions & $p_{\lambda_1}=\sum_{j=1}^m x_j^{\lambda_1}$ & $p_{\lambda}=\prod_{i=1}^l p_{\lambda_i}$ \\ \hline
elementary\index{symmetric functions!elementary} functions & $e_{\lambda_1} =\sum_{\small{j_1 <j_2 < \ldots < j_{\lambda_1}}} x_{j_1} \ldots x_{j_{\lambda_1}}$ & $e_{\lambda} = \prod_{i=1}^l e_{\lambda_i}$ \\ \hline
complete homogeneous\index{symmetric functions!complete homogeneous} functions & $h_{\lambda_1} =\sum_{\small{j_1 \leq j_2 \leq \ldots \leq j_{\lambda_1}}} x_{j_1} \ldots x_{j_{\lambda_1}}$ & $h_{\lambda} = \prod_{i=1}^l h_{\lambda_i}$ \\ \hline
\end{tabular}
\end{center}

\vspace{.25cm}

Another important basis is given by the monomial\index{symmetric functions!monomial} functions $m$,
\[
m_{\lambda} = \sum_{\mbox{\small{$\sigma \in S_{\lambda}$}}} x_{\sigma(1)}^{\lambda_1} x_{\sigma(2)}^{\lambda_2} \ldots x_{\sigma(m)}^{\lambda_m}~;
\]
here $S_{\lambda}$ is the set of permutations giving distinct terms in the sum; $\lambda$ is considered as infinite.

The last basis we mentioned distinguishes itself from the other ones in two ways; the advantage is that it is very easy to visualize, and proving that it is indeed a basis is immediate. The disadvantage is that it is not multiplicative\footnote{For $x \in \{p,e,h\}$, $x_{\lambda} x_{\mu} = x_{\rho}$, where $\rho$ can be obtained in an algorithmic fashion from $\lambda$ and $\mu$ (sometimes by mere concatenation and reordering). In general, $m_{\lambda}m_{\mu}$ is not a monomial.}.

Monomial\index{symmetric functions!monomial}s seem to be the basis of choice for most people working in statistics or engineering. Combinatorialists often prefer to express series in the power-sum\index{symmetric functions!power-sum} basis, because of connections with character theory.

%\subsection{Multivariate Orthogonal Polynomials}

\subsection{Multivariate Gamma Function} \label{gammafunc}

Before we proceed, we will need to define the multivariate Gamma
function for arbitrary $\alpha$; the
real and complex versions are familiar from the literature, and the
arbitrary $\alpha>0$ case represents an immediate extension:
\begin{eqnarray} \label{Gamma_def}
\Gamma^{\alpha}_m (a) ~~~ = ~~~ \pi^{m(m-1)/(2\alpha)} ~~ \prod_{i=1}^m
~\Gamma \left(a-\frac{i-1}{\alpha} \right)~.
\end{eqnarray}
                                                                                
Just as the univariate Gamma function generalizes to the multivariate one,
the shifted factorial (Pochhammer
symbol, rising
factorial) 
\[(a)_{k} =
\frac{\Gamma(a+k)}{\Gamma(a)}\] becomes the generalized shifted factorial.                                                               
We call
\begin{eqnarray} \label{poch_def}
(a)_{\kappa}^{\alpha} ~=~ \prod_{i=1}^{length(\kappa)}
\left(a-\frac{i-1}{\alpha}\right)_{\kappa_i} ~=~
\prod_{i=1}^{length(\kappa)} \frac{\Gamma \left(a- \frac{i-1}{\alpha}
+\kappa_i \right)}{\Gamma \left(a-\frac{i-1}{\alpha}\right)}~
\end{eqnarray}
the \emph{generalized shifted factorial}, or \emph{generalized Pochhammer
symbol}.

\subsection{Jack Polynomials (the Multivariate Monomials)}

%The Jack polynomials\index{Jack polynomials} $C_{\lambda}^{\alpha}$ constitute a far more complex class of homogeneous bases (depending on the parameter $\alpha$) than any of the previously mentioned ones. 

Jack polynomials allow for several equivalent definitions (up to certain normalization constraints). In addition to the definition presented in the introduction (at the end of Section \ref{mm}), we present here two more (Definitions \ref{oo} and \ref{tt}). Definition \ref{oo} arose in combinatorics, whereas Definition \ref{tt} arose in statistics. We will mainly work with Definition \ref{tt}. 

\begin{definition} \label{oo}(following Macdonald\index{Macdonald, I.} \cite{MacDonald_book}) The Jack polynomials\index{Jack polynomials!combinatorial definition} $P_{\lambda}^{\alpha}$ are orthogonal with respect to the inner product\index{Jack polynomials!inner product} defined below on power-sum\index{symmetric functions!power-sum} functions 
\[
\langle p_{\lambda}, p_{\mu} \rangle_{\alpha} = \alpha^{l(\lambda)} z_{\lambda} \delta_{\lambda \mu},
\]
where $z_{\lambda} = \prod\limits_{i=1}^{l(\lambda)} a_i!i^{a_i}$,  $a_i$ being the number of occurrences of $i$ in $\lambda$. In addition, 
\[
P_{\lambda}^{\alpha} = m_{\lambda} + \sum_{\mu \preceq \lambda} u_{\lambda, \mu}^{\alpha} m_{\mu}~.
\]
\end{definition}

There are two main normalizations of the Jack polynomials\index{Jack polynomials!normalizations} used in combinatorics, the ``J'' normalization (which makes the coefficient of the lowest-order monomial\index{symmetric functions!monomial}, $[1^n]$, be exactly $n!$) and the ``P'' normalization (which is monic, and is given in Definition \ref{oo}). To convert between these normalizations, see Tables \ref{ident} and \ref{conversions}. In Table \ref{ident}, $I_m = (1,1,1,\ldots, 1)$, where the number of variables is $m$. 

We use the notation $\kappa \vdash k$ for $\kappa$ a partition\index{partition} of $k$, and $\rho_{\kappa}^{\alpha}$ for $\sum_{i=1}^m k_i(k_i - 1 - \frac{2}{\alpha} (i-1))$.

\begin{definition} \label{tt}(following Muirhead\index{Muirhead, R.}, \cite{muirhead82a}) The Jack polynomial\index{Jack polynomials!eigenfunction definition} $C_{\kappa}^{\alpha}$ is the only homogeneous polynomial eigenfunction of the following Laplace\index{Laplace-Beltrami-type operators}-Beltrami-type operator
\begin{eqnarray*} 
D^{*}= \sum_{i=1}^m x_i^2 \frac{d^2}{dx_i^2} + \frac{2}{\alpha} \sum_{1 \leq i \neq j \leq m } \frac{x_i^2}{x_i-x_j} \frac{d}{dx_i}~,
\end{eqnarray*}
with eigenvalue $\rho_{\kappa}^{\alpha}+k(m-1)$, having highest-order term corresponding to $\kappa$. In addition, 
\[
\sum_{\small{\kappa ~\vdash ~k,~~l(\kappa) \leq m}} C_{\kappa}^{\alpha}(x_1, x_2,\ldots, x_m) = (x_1+x_2+\ldots x_m)^k~.
\]
\end{definition}

\begin{remark}
The ``C'' normalization for the Jack polynomial allows for defining scalar hypergeometric function\index{hypergeometric functions!definition}s of multivariate (or matrix) argument. These are useful for computing Selberg\index{Selberg integral}-type integrals and other quantities which appear in various fields, from the theory of random walks to multivariate statistics and quantum many-body problems. 
\end{remark}

\begin{remark} David M. Jackson\index{Jackson, D.M.} \cite{jackson_pers} pointed out that the $D^*$ operator also appears in algebraic geometry, for example in the context of ramified covers. 
\end{remark}

\begin{definition} 
Given the diagram of a partition\index{partition} $\kappa$ (see Figure 1), define $a_{\kappa}(s)$ (the ``arm\index{arm-length!definition}-length'') as the number of squares to the right of $s$; $l_{\kappa}(s)$ (the ``leg-length\index{leg-length!definition}'') as the number of squares below $s$; $h_{\kappa}^{*}(s) =  l_{\kappa}(s) + \alpha(1+a_{\kappa}(s))$ (the ``upper hook length'') and $h_{*}^{\kappa}(s) = l_{\kappa}(s) + 1+ \alpha a_{\kappa}(s)$ (the ``lower hook length'').
\end{definition}

\begin{figure}[!ht]
 \caption{   
 \noindent The Arm-length and the Leg-length.}
 \begin{center}
 \psfig{figure=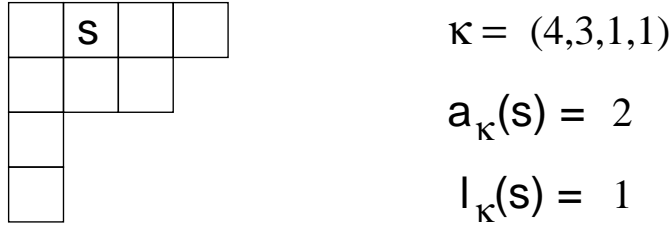, height = 3cm}
 \end{center}
\end{figure}
\index{arm-length!definition}\index{leg-length!definition}

Finally, a further definition is needed in order to present the conversion table.

\begin{definition} 
Let
\begin{eqnarray*}
c(\kappa, \alpha) & = &  \prod\limits_{s \in \kappa} h^*_{\kappa} (s)~, \\
c'(\alpha, \kappa) & = & \prod\limits_{s \in \kappa} h_{*}^{\kappa} (s)~,\\
j_{\kappa} & = & c(\alpha, \kappa)~ c'(\alpha, \kappa)~,
\end{eqnarray*}
where $h^*_{\kappa}$ and $h_*^{\kappa}$ have been defined above.
\end{definition}
\index{Jack polynomials!conversion between normalizations}

To explain the conversions between ``J'', ``P'', and ``C'', we recall the definition of the generalized Gamma function and generalized shifted factorial from Section \ref{gammafunc}.

We can now present Tables \ref{ident} and \ref{conversions}; the entries have been filled out using James\index{James, A.T.} \cite{james68}, Forrester\index{Forrester, P.J.} and Baker\index{Baker, T.} \cite{Forrester_poly}, and Stanley\index{Stanley, R.P.} \cite{stanley_jacks}.

\begin{table}[ht]
\caption{Values of the different normalizations of Jack polynomials of partition\index{partition} $\kappa$ and parameter $\alpha$ at $I_m$.}
\begin{center}
\begin{tabular}{|c|c|c|} \hline
Normalization & Basic Property & Value at $I_m = (1,1,1,\ldots, 1)$ \\ 
\hline
C & sums to $(x_1 +x_2+\ldots+x_n)^k$  & $~~~~C_{\kappa}^{\alpha} (I_m) = \frac{\alpha^{2k} k!}{j_{\kappa}} \Big(\frac{m}{\alpha} \Big)_{\!\kappa}~$ \\ \hline
J & has trailing coefficient $n!$ & $~J_{\kappa}^{\alpha} (I_m) = \alpha^k \Big(\frac{m}{\alpha} \Big)_{\!\kappa}~$ \\ \hline
P & is monic & $~~~~~P_{\kappa}^{\alpha} (I_m) = \frac{\alpha^{k}}{c(\alpha, \kappa)} \Big(\frac{m}{\alpha} \Big)_{\!\kappa}~$ \\ \hline
\end{tabular}
\label{ident}
\end{center}
\end{table}

\begin{table}[ht!]
\caption{Conversions between the three normalizations for the Jack polynomials; the $a(V,W)$ entry above is defined as 
$V_{\kappa}^{\alpha}(x_1, \ldots, x_m)$ $=$$ a(V,W) W_{\kappa}^{\alpha}(x_1, \ldots, x_m)$.}
\begin{center}
\begin{tabular}{|c|c|c|c|} \hline
& C & J & P \\ \hline
& & & \\
C &  & $\frac{\alpha^k~k!}{j_k}$ & $\frac{\alpha^k~k!}{c'(\kappa, \alpha)}$  \\ & & & \\ \hline
& & & \\
J & $\frac{j_k}{\alpha^k~k!}$ & & $c(\kappa, \alpha)$\\ & & & \\ \hline
& & & \\
P & $\frac{c'(\kappa, \alpha)}{\alpha^k~k!}$ & $\frac{1}{c(\kappa, \alpha)}$ & \\ & & & \\ \hline
\end{tabular}
\label{conversions}
\end{center}
\end{table}

\subsection{Algorithm used to compute the Jack Polynomials} \label{jacks_comp}

From the Laplace-Beltrami equation, one can find an expansion for the Jack polynomials\index{Jack polynomials!algorithm} of the type
\[
C_{\kappa}^{\alpha} (x_1, x_2, \ldots, x_m) = \sum_{\lambda \leq \kappa} c_{\kappa, \mu}^{\alpha} m_{\lambda}(x_1,x_2,\ldots,x_m)~,
\]
where $\lambda$ and $\kappa$ are both partitions of the same integer $|\kappa|$, and the order imposed on partitions is the lexicographic one. The coefficients $c_{\kappa, \lambda}^{\alpha}$ depend on all three parameters; $m_{\lambda}(x_1,x_2, \ldots, x_m)$ is the monomial function corresponding to $\lambda$. 

Note that as a consequence of the above, if $l(\kappa)>m$, $C_{\kappa}^{\alpha} (x_1, x_2, \ldots,x_m)= 0$ (``there is no highest-order term'').  

Using the eigenfunction equation 
\begin{eqnarray} \label{jack_eig_eq}
D^{*} C_{\kappa}^{\alpha} = (\rho_{\kappa}^{\alpha} + k(m-1))C_{\kappa}^{\alpha}~,
\end{eqnarray}
where 
\[
\rho_{\kappa}^{\alpha} = \sum_{i=1}^m k_i(k_i - 1 - \frac{2}{\alpha} (i-1))~
\]
one can obtain a recurrence for $c_{\kappa, \lambda}^{\alpha}$ from which the Jack polynomials\index{Jack polynomials!algorithm} can be explicitly calculated. This recurrence is
\begin{eqnarray} \label{recur_jack}
c_{\kappa, \lambda}^{\alpha} = \frac{\frac{2}{\alpha}}{\rho_{\kappa}^{\alpha} - \rho_{\lambda}^{\alpha}} \sum_{\lambda <\mu \leq \kappa} \Big((l_i+t)- (l_j - t) \Big) c_{\kappa, \mu}^{\alpha}~,
\end{eqnarray}
where $\lambda = (l_1, \ldots, l_i, \ldots, l_j, \ldots, l_m)$, $\mu = (l_1, \ldots, l_i+t, \ldots, l_j-t, \ldots, l_m)$, and $\mu$ has the property that, when properly reordered, it is between $\lambda$ (strictly) and $\kappa$ in lexicographic order.

In fact we can do better, using two propositions found in Macdonald\index{Macdonald, I.}'s book \cite[(10.13), (10.15)]{MacDonald_book}. Roughly the content of the two propositions is that the Jack polynomials\index{Jack polynomials!algorithm}, in ``P'' normalization, can be written as
\[
P_{\kappa}^{\alpha} = m_{\kappa} + \sum_{\lambda \prec \kappa} u_{\kappa, \lambda}^{\alpha} m_{\lambda}~,\]
with $ u_{\kappa, \lambda}^{\alpha} >0$ whenever $\kappa \succ \lambda$ (the ordering imposed on partitions here is the dominance ordering).

Thus it follows that the recurrence can be improved to 
\begin{eqnarray} \label{recurence_jack}
c_{\kappa, \lambda}^{\alpha} = \frac{\frac{2}{\alpha}}{\rho_{\kappa}^{\alpha} - \rho_{\lambda}^{\alpha}} \sum_{\lambda \prec \mu \preceq \kappa} \Big((l_i+t)- (l_j - t) \Big) c_{\kappa, \mu}^{\alpha}~,
\end{eqnarray}
where $\lambda = (l_1, \ldots, l_i, \ldots, l_j, \ldots, l_m)$, $\mu = (l_1, \ldots, l_i+t, \ldots, l_j-t, \ldots, l_m)$, and $\mu$ has the property that, when properly reordered, it is between $\lambda$ (strictly) and $\kappa$ in domination order.

This recurrence, at first glance, seems to be enough to compute all coefficients $c_{\kappa, \lambda}^{\alpha}$, once $c_{\kappa, \kappa}^{\alpha}$ is found. However, one has to account for the possibility that $\rho_{\kappa}^{\alpha} = \rho_{\lambda}^{\alpha}$ for some $\lambda$ different from $\kappa$; what can one do in that case? 

Fortunately, this never happens. We first need the following well known Proposition.

\begin{proposition} \label{well_known}
The dominance ordering is a lattice on the set of partitions of a given number. In particular, between any partitions $\kappa$ and $\lambda$ such that $\kappa \succ \lambda$, there exists a ``path'' on this lattice, $\sigma^0 = \kappa \succ \sigma^1 \succ \ldots \succ \sigma^t = \lambda$, such that $\sigma^{i+1}$ differs from $\sigma^i$ in the following way: there exists $i_1<i_2$ such that $\sigma^{i+1}$ and $\sigma^{i}$ agree in all places but $i_1$ and $i_2$, $(\sigma^{i+1})_{i_1} = (\sigma^{i})_{i_1}-1$, and $(\sigma^{i+1})_{i_2} = (\sigma^{i})_{i_2}+1$.
\end{proposition}

Now we can prove that we never divide by $0$ in computing Recurrence \ref{recurence_jack}.

\begin{lemma} \label{no_dividing_by_0}
If $\lambda \prec \kappa$, then $\rho_{\lambda}^{\alpha} \neq \rho_{\kappa}^{\alpha}$, for all $\alpha>0$.
\end{lemma}

\begin{proof} Let $\lambda \prec \kappa$ be two partitions, let $m = \max \{length(\kappa), length(\lambda) \}$ , and assume that there is some $\alpha>0$ such that 
\[
\rho_{\lambda}^{\alpha} = \rho_{\kappa}^{\alpha}~.
\]
Since the two partitions sum to the same number, the above is equivalent to 
\[
\sum_{i=1}^m k_i^2 - \lambda_i^2 = \frac{2}{\alpha} \sum_{i=1}^m (k_i - \lambda_i)(i-1)~.
\]
The right-hand side is non-negative (as an immediate consequence of the strict ordering).

We show that the left-hand side is positive by induction. For that we will use Proposition \ref{well_known}, which shows that it is enough to prove that 
\[
\sum_{i=1}^m k_{i}^2 - \lambda_{i}^2 \geq 0~
\]
in the case when $\kappa$ and $\lambda$ differ only in two places, $i_1<i_2$. Note that if $\kappa_{i_1} = \lambda_{i_1}+1$ and $\kappa_{i_2} = \lambda_{i_2}-1$, this implies that $\kappa_{i_1} \geq \kappa_{i_2}+2$. Hence 
\[
\sum_{i=1}^m k_{i}^2 - \lambda_{i}^2 = k_{i_1}^2 - \lambda_{i_1}^2+ k_{i_2}^2 - \lambda_{i_2}^2 = 2 k_{i_1}-1 - 2k_{i_2} -1\geq 2 >0~,
\]
and we are done.
\end{proof}

Proposition \ref{well_known} ensures thus that once $c_{\kappa \kappa}^{\alpha}$ is determined, every other non-zero coefficient is uniquely determined. 

Finally, for $c_{\kappa \kappa}^{\alpha}$ we use the following formula (deduced on the basis of Table \ref{ident} and the fact that $P_{\kappa}^{\alpha}$ has highest-order coefficient $1$):
\[
c_{\kappa \kappa}^{\alpha} = \frac{\alpha^{k} k!}{c'(\kappa, \alpha)}~.
\]

\begin{remark} It is worth mentioning that, from the recurrence \ref{recurence_jack}, by letting $\alpha \rightarrow \infty$, the coefficient $c_{\kappa, \lambda}^{\alpha}$ goes to $0$ faster than $c_{\kappa, \kappa}^{\alpha}$, for any $\lambda \neq \kappa$. Thus, at $\alpha = \infty$, the Jack ``P'' polynomial (which is monic) is the symmetric monomial. This could also be seen from the weight functions, as at $\alpha = \infty$, the ``interdependence'' term $\prod\limits_{1\leq i<j \leq n}|x_i - x_j|^{2/\alpha}$ (see for example \ref{general_weight}) disappears and the variables separate. \end{remark}

\subsection{Multivariate binomial coefficients} \label{gbc}

%These correspond to the ``shifted Jack polynomials'' which first appear in general $\alpha$ case in the work of Okounkov (reference); however, for the particular cases $\alpha = 1$ and $2$ they had been discovered and studied by statisticians in the '60s (James, Constantine, Muirhead). 

Many algebraic quantities (and the identities they satisfy)  can be extended from the univariate case to the multivariate case through Jack polynomials. One such example is the multivariate, or generalized, binomial coefficient.

\begin{definition}\index{generalized binomial coefficients!definition}
We define the multivariate (or generalized) binomial coefficients ${\kappa \choose \sigma}$ as
\[
\frac{C_{\kappa}^{\alpha}(x_1+1, x_2+1, \ldots, x_m+1)}{C_{\kappa}^{\alpha}(1,1,\ldots, 1)} = \sum_{s=0}^{k} \sum_{\small{ \sigma \vdash s,~\sigma \subseteq \kappa}} {\kappa \choose \sigma} \frac{C_{\sigma}^{\alpha}(x_1, x_2, \ldots, x_m)}{C_{\sigma}^{\alpha}(1,1,\ldots,1)}~,
\]
where $\sigma \subset \kappa$ means that $\sigma_i \leq \kappa_i$ for all $i$.
\end{definition}

%These correspond to the ``shifted Jack polynomials'' which first appear in general $\alpha$ case in the work of Okounkov (reference); however, for the particular cases $\alpha = 1$ and $2$ they had been discovered and studied by statisticians in the '60s (James\index{James, A.T.}, Constantine, Muirhead\index{Muirhead, R.}). 
\index{generalized binomial coefficients}
The generalized binomial coefficients depend on $\alpha$, but are independent of both the number of variables $m$ and the normalization of the Jack polynomials (the latter independence is easily seen from the definition).

The multivariate binomial coefficients\index{multivariate binomial coefficients!see{ generalized binomial coefficients}} generalize the univariate ones; some simple properties of the former are straightforward generalizations of properties of the latter. For example, 
\begin{eqnarray*}
& {\kappa \choose (0)} & =  1~,\\
&{\kappa \choose (1)} & =  |\kappa| ~,\\
&{\kappa \choose \sigma}& = 0 ~\mbox{if}~ \sigma \not \subseteq \kappa~,\\
&{\kappa \choose \sigma} & = \delta_{\kappa} ~\mbox{if}~ |\kappa|=|\sigma|~, \\
&{\kappa \choose
 \sigma} & \neq  0 ~\mbox{if}~ |\kappa| = |\sigma|+1,~\mbox{iff} ~\sigma = \kappa_{(i)}~,
\end{eqnarray*}
where $\kappa_{(i)} = (k_1, \ldots, k_i-1, \ldots, k_m)$. The above are true for all $\kappa$ and $\alpha$, and $\sigma$ subject to the constraints.

\subsection{Algorithm used to compute the multivariate binomial coefficients}

One can prove, using the eigenfunction equation (\ref{jack_eig_eq}) and the definition of the generalized binomial coefficients, that 
\begin{eqnarray} \label{recur_gbinom}
\sum_i {\sigma^{(i)} \choose \sigma} {\kappa \choose \sigma^{(i)}} = (k-s) {\kappa \choose \sigma}~,
\end{eqnarray}
where $|\sigma| = s,~|\kappa|=k$, $\sigma^{(i)} = (\sigma_1 \ldots, \sigma_i+1, \ldots, \sigma_m)$. All generalized binomial coefficients can be found recursively, once one has a way to compute the so-called ``contiguous'' coefficients ${\sigma^{(i)} \choose \sigma}$. 

To compute the contiguous coefficients, we use Proposition 2 from \cite{kaneko}, applied to $\kappa= \sigma^{(i)}$, and simplified slightly:
\begin{eqnarray} \label{cont_binom}
{\sigma^{(i)} \choose \sigma } = j_{\sigma}^{-1} g_{\sigma~1}^{\sigma^{(i)}}~,
\end{eqnarray}
where $g_{\sigma~1}^{\sigma^{(i)}}$ is 
\[
g_{\sigma~1}^{\sigma^{(i)}} = \left( \prod_{s \in \sigma} A_{\sigma^{(i)}} \right)\left( \prod_{s \in \sigma} B_{\sigma^{(i)}} \right)~.
\]
Here
\begin{eqnarray*}
A_{\sigma^{(i)}} & = & \left \{ \begin{array}{l} h_{*}^{\sigma}(s),~~~\mbox{if $s$ is not in the $i$th column of $\sigma$}~, \\  h^{*}_{\sigma}(s),~~~\mbox{otherwise}. \end{array} \right . \\
B_{\sigma^{(i)}} & = & \left \{ \begin{array}{l} h^{*}_{\sigma^{(i)}}(s),~~~\mbox{if $s$ is not in the $i$th column of $\sigma$}~, \\  h_{*}^{\sigma^{(i)}}(s),~~~\mbox{otherwise}~. \end{array} \right . 
\end{eqnarray*}
Knowing the contiguous coefficients allows for computing all the generalized binomial coefficients. 

\begin{remark}
The generalized binomial coefficients are independent of the number of variables. They are rational functions of $\alpha$.
\end{remark}

\subsection{Multivariate Orthogonal Polynomials} \label{orthopol}

%In this section we finally define three types of multivariate orthogonal polynomials, namely Hermite, Laguerre, and Jacobi, and we explain how MOPs computes them.

%To refresh the reader's memory, we recall the univariate definition of a set of orthogonal polynomials $P_{i}(x)$, $i=0, 1, \ldots$ with respect to a weight function $w(x)$ supported on an interval $\Omega \in \mathbb{R}$:
%\[
%\int_{\Omega} P_i(x)~P_j(x)~w(x)~\mbox{d}x ~ = ~ c_{i,j}~\delta_{ij}, ~~~\forall i,j ~\in \mathbb{N}~.
%\]
%where the constant $c_{i,j}$ depends on the normalization of the set of polynomials $P_i(x)$.

%One can think of the polynomials $P_i(x)$ are being the result of a Gram-Schmidt orthogonalization process with the inner product
%\[
%<f,~g> ~ = ~ \int_{\Omega} f(x)~g(x)~w(x)~\mbox{d}x~,
%\]
%assuming that one starts with the monomial basis $x^i$, $i=0,1, \ldots$.

%The multivariate definition for orthogonal polynomials is quite similar, except that this time $\Omega \in \mathbb{R}^n$, $w(x)$ is a scalar function of a vector, and the indexing is over partitions $\lambda$:
%\[
%\int_{\Omega} P_{\lambda}(x) ~P_{\kappa}(x) ~w(x) )~\mbox{d}x ~ = ~ c_{\kappa, \lambda}~\delta_{\kappa \lambda}, ~~~\forall \lambda,~\kappa ~\mbox{finite partitions}~.
%\]

%Again, $c_{\kappa, \lambda}$ depends on the normalization of the set of polynommials $P_{\lambda}(x)$.

\subsubsection{Jacobi Polynomials}

These polynomials represent the Gram-Schmidt orthogonalization of the Jack polynomials $C_{\lambda}^{\alpha}$ with respect to the Jacobi weight function
\begin{eqnarray} \label{jac_density}
d\mu_{J}^{\alpha}(x_1, \ldots, x_m)& =& \frac{\Gamma_m^{\alpha} \big(g_1+g_2+\frac{2}{\alpha}(m-1)+2 \big)}{\Gamma_m^{\alpha} (g_1+\frac{m-1}{\alpha}+1 \big)} \prod_{i=1}^m \Big[ x_i^{g_1} (1-x_i)^{g_2} \Big] ~~\times \\
 & & ~~~~~~~~~\times ~~\prod_{i <j} |x_i - x_j|^{2/\alpha} dx_1 \ldots dx_m~,
\end{eqnarray}
on the hypercube $[0,1]^m$. For the purpose of well-definitedness we assume 
\begin{eqnarray} \label{bound_c}
g_1, g_2> -1~.
\end{eqnarray}

Define
\begin{eqnarray*}
\delta^{*} & = & \sum_i x_i \frac{d^2}{dx_i^2} + \frac{2}{\alpha} \sum_{i \neq j} \frac{x_i}{x_i-x_j} \frac{d}{dx_i}~\\
E & = & \sum_i x_i \frac{d}{dx_i} ~\\
\epsilon  & = & \sum_i \frac{d}{dx_i}~;
\end{eqnarray*}
then the Jacobi polynomials are eigenfunctions of the following Laplace-Beltrami operator:
\begin{eqnarray} \label{jacobi_op}
D^{*} +(g_1+g_2+2)E -\delta^{*} -(g_1+1)\epsilon~,
\end{eqnarray}
with eigenvalue $\rho_{\kappa}^{\alpha} +|\kappa|(g_1+g_2+\frac{2}{\alpha}(m-1)+2)$. 

%Let $g_1 = a - \frac{m-1}{\alpha} -1$, $g_2 = c - a -  \frac{m-1}{\alpha} -1$.

\subsubsection{Algorithm used to compute the Jacobi Polynomials} \label{format_jacobi}

Using the fact that the Jacobi polynomials are eigenfunctions of the operator (\ref{jacobi_op}), one obtains that these polynomials can be written in Jack polynomial basis as
\[
J _{\kappa}^{\alpha, g_1, g_2} (x_1, \ldots, x_m) = 
(g_1+\frac{m-1}{\alpha}+1)_{\kappa} C_{\kappa}^{\alpha}(I_m) \sum_{\sigma 
\subseteq \kappa} \frac{(-1)^{s} c_{\kappa 
\sigma}^{\alpha}}{(g_1+\frac{m-1}{\alpha}+1)_{\sigma}}  
\frac{C_{\sigma}^{\alpha}(x_1, \ldots, x_m)}{C_{\sigma}^{\alpha}(I_m)}~,
\]
where the coefficients $c_{\kappa \sigma}^{\alpha}$ satisfy the recurrence
\begin{eqnarray} \label{recurrence_jacobi}
c_{\kappa \sigma}^{\alpha} = \frac{1}{ \Big((g_2+g_1+\frac{2}{\alpha}(m-1)+2)(k-s)+ \rho_{\kappa}^{\alpha} - \rho_{\sigma}^{\alpha} \Big)} \sum_{\small{i~\mbox{allowable}}} {\kappa \choose \sigma^{(i)}} {\sigma^{(i)} \choose \sigma} c_{\kappa \sigma^{(i)}}^{\alpha}~,
\end{eqnarray}
with the previous notation for $\rho_{\kappa}^{\alpha}$ and $\sigma^{(i)}$. The question is again whether the denominator is always nonzero.

\begin{proposition}
Under these assumptions, $(g_2+g_1+\frac{2}{\alpha}(m-1)+2)(k-s)+ \rho_{\kappa}^{\alpha} - \rho_{\sigma}^{\alpha}$ is never $0$. 
\end{proposition}

\begin{proof}
The proof is very similar with the corresponding proof of Section \ref{jacks_comp}; the two crucial facts here are that one needs one show it for the case $\kappa = \sigma^{(i)}$, and that $g_1$ and $g_2$ are both larger than $-1$ (due to (\ref{bound_c})).
\end{proof}

Letting $c_{\kappa, \kappa}^{\alpha}=1$ for all $\kappa$ and $\alpha$ allows all the coefficients to be uniquely determined.

\subsubsection{Laguerre Polynomials}

The multivariate Laguerre polynomials are orthogonal with respect to the Laguerre weight function 
\begin{eqnarray} \label{lag_density}
d\mu_{L}^{\alpha} (x_1, \ldots, x_m) = \frac{1}{\Gamma_m^{\alpha} (\gamma+\frac{m-1}{\alpha}+1)} e^{-\sum_i x_i} \prod_i x_i^{\gamma} \prod_{i \neq j} |x_i - x_j|^{2/\alpha} dx_1 \ldots dx_m~,
\end{eqnarray}
on the interval $[0, \infty)^m$. Note that for the purpose of well-definitedness, we must have $\gamma>-1$.

This weight function can be obtained from the Jacobi weight function (\ref{jac_density}) of the previous subsection by substituting $(g_1+g_2+\frac{2}{\alpha}(m-1)+2)^{-1}(x_1, \ldots, x_m)$ for $(x_1, \ldots, x_m)$ and then taking the limit as $ g_2 \rightarrow \infty$. The same limiting process applied to the Jacobi polynomials yields the Laguerre polynomials.

Under the transformation mentioned above, the Jacobi differential operator becomes
\begin{eqnarray} \label{laguerre_op}
\delta^{*} -E+(\gamma+1) \epsilon~,
\end{eqnarray}
and the Laguerre polynomials are eigenfunctions of this operator with eigenvalue $|\kappa|$. 

\subsubsection{Algorithm used to compute the Laguerre Polynomials} \label{format_laguerre}

The Laguerre polynomials have an explicit expansion in Jack polynomial basis, which depends on the generalized binomial coefficients:
\[
L_{\kappa}^{\alpha, \gamma} (x_1, \ldots, x_m)= 
\small{\mbox{$(\gamma+\frac{m-1}{\alpha}+1)$}}_{\kappa} 
C_{\kappa}^{\alpha}(I_m) \sum_{\sigma \subseteq \kappa} \frac{(-1)^s {\kappa \choose \sigma}}{\small{\mbox{$(\gamma+\frac{m-1}{\alpha}+1)$}}_{\sigma}} \frac{C_{\sigma}^{\alpha}(x_1, \ldots, x_m)}{C_{\sigma}^{\alpha}(I_m)}~.
\]

Note that the coefficient of $C_{\kappa}^{\alpha}(x_1, \ldots, x_m)$ in $L_{\kappa}^{\alpha, \gamma} (x_1, \ldots, x_m)$ is $(-1)^k$.

\subsubsection{Hermite Polynomials}

The multivariate Hermite polynomials are orthogonal with respect to the Hermite weight function 
\begin{eqnarray} \label{herm_density}
d\mu_{H}^{\alpha}(x_1,\ldots,x_m) & = & 2^{-m/2}~\pi^{m(m-1)/\alpha-m/2}~\frac{(\Gamma(1+\frac{1}{\alpha}))^m}{\Gamma_m^{\alpha}(1+\frac{m}{\alpha})} ~~\times \\
& & ~~~~~~~~~\times ~~ ~e^{-\sum_{i=1}^m x_i^2/2} \prod_{i \neq j} |x_i - x_j|^{2/\alpha} dx_1 \ldots dx_m~,
\end{eqnarray}
on $\mathbb{R}^m$.

This weight function can be obtained by taking $(\gamma+\sqrt{\gamma}x_1, \gamma+\sqrt{\gamma}x_2, \ldots, \gamma+\sqrt{\gamma}x_m)$ in (\ref{lag_density}), and then letting $\gamma$ go to infinity; note that the only remaining parameter is $\alpha$. 

Under this limiting process, the differential operator becomes 
\begin{eqnarray} \label{herm_op}
\delta^{**} - E~,
\end{eqnarray}
where 
\[
\delta^{**} = \sum_i \frac{d^2}{dx_i^2} + \frac{2}{\alpha} \sum_{i \neq j} \frac{1}{x_i - x_j} \frac{d}{dx_i}~.
\]

The Hermite polynomials\index{multivariate Hermite polynomials!definition} are eigenfunctions of this operator with eigenvalue $|\kappa|$. 

\begin{remark} \label{herm_from_lag} Similarly, 
\begin{eqnarray} 
\lim_{\gamma \rightarrow \infty} \gamma^{-k/2} L^{\alpha, \gamma}_{\kappa}( \gamma+\sqrt{\gamma}x_1, \gamma+\sqrt{\gamma}x_2, \ldots, \gamma+\sqrt{\gamma}x_m) = (-1)^k H_{\kappa}^{\alpha}(x_1, \ldots, x_m)~.
\end{eqnarray}
\end{remark}

\subsubsection{Algorithm used to compute the Hermite Polynomials} \label{format_hermite}

Using the corresponding Hermite differential operator (\ref{herm_op}), we obtain the following recurrence for the coefficients of the polynomial. Let 
\[
H_{\kappa}^{\alpha}(x_1, \ldots, x_m) = \sum_{\sigma \subseteq \kappa} c_{\kappa, \sigma}^{\alpha} \frac{C_{\sigma}^{\alpha}(x_1, \ldots, x_m)}{C_{\sigma}(I_m)}~;
\]
then
\begin{eqnarray} \label{recurrence_hermite}
c_{\kappa, \sigma}^{\alpha} = \frac{1}{k-s} \left( \sum_{i} {\sigma^{(i)(i)} \choose \sigma^{(i)}} {\sigma^{(i)} \choose \sigma} c_{\kappa, \sigma^{(i)(i)}}^{\alpha} + \sum_{i<j} (\sigma_i -\sigma_j - \frac{1}{\alpha}(i-j)) {\sigma^{(i)(j)} \choose \sigma^{(j)}} {\sigma^{(j)} \choose \sigma} c_{\kappa, \sigma^{(i)(j)}}^{\alpha} \right).
\end{eqnarray}

In the above, $i<j$ take on all admissible values, and we choose $c_{\kappa, \kappa}^{\alpha} = C_{\kappa}^{\alpha}(I_m)$.

Alternatively, we can obtain the coefficients directly through the limiting process described in Remark \ref{herm_from_lag}:

\begin{eqnarray} \label{mine}
\Big[ C_{\sigma}^{\alpha}(X) \Big] H_{\kappa}^{\alpha}(X)~ =~ (-1)^k
\frac{C_{\kappa}^{\alpha}(I_m)}{C_{\sigma}^{\alpha}(I_m)} 
\sum_{j=s}^{\frac{k+s}{2}} (-1)^{k-j} \sum_{\sigma \subseteq \mu \subseteq 
\kappa; \mu \vdash j} {\kappa \choose \mu} {\mu \choose \sigma} \Big[ 
r^{\frac{k+s}{2} - j} \Big] F(r, \alpha, m, \kappa, \sigma)~,
\end{eqnarray}
where\[
F(r, \alpha, m, \kappa, \sigma) ~ = ~ \frac{(r+\frac{1}{\alpha}(m+\alpha-1))_{\kappa}}{(r+\frac{1}{\alpha}(m+\alpha-1))_{\sigma}}~.
\]

We use the above formula to calculate a single coefficient, $c_{\kappa, []}^{\alpha}$, for reasons of smaller computational complexity (in computing integrals with respect to the Hermite weight function; see Section \ref{integr}).

%Here $\Big[ \ldots \Big]$ is the notation for a coefficient in a polynomial/power series.

Note that if $\sigma \not \subseteq \kappa$ or $k \neq s\pmod{2}$, then the above is $0$.

%To the best of our knowledge, equation (\ref{mine}) is new and it does not apear in the literature. It can be derived without much effort by applying the limiting process described above to the Jack polynomial expansion of the corresponding Laguerre polynomial.

\subsection{Hypergeometric functions}

The hypergeometric function\index{hypergeometric functions!definition}s are perhaps the easiest to generalize from univariate to multivariate. For the multivariate versions, a good reference is Forrester\index{Forrester, P.J.}'s unpublished book \cite{Forrester_book}.

\begin{definition}
We define the hypergeometric function\index{hypergeometric functions!definition} ${}_pF_{q}^{\alpha}$ of parameters $a_1, \ldots, a_p$, respectively $b_1, \ldots, b_q$ and of variables $(x_1, \ldots, x_m)$ by
\[
{}_pF_{q}^{\alpha}(a_1, \ldots, a_p; b_1, \ldots, b_q; x_1, \ldots, x_m) = \sum_{k=0}^{\infty} \sum_{\kappa \vdash k} ~\frac{(a_1)_{\kappa} \ldots (a_p)_{\kappa}}{k!~(b_1)_{\kappa} \ldots (b_q)_{\kappa}} ~C_{\kappa}^{\alpha}(x_1, \ldots, x_m)~.
\]
\end{definition}

Note that this is a formal definition; $p \leq q$ is needed in order for
the hypergeometric series to converge everywhere, and when $p=q+1$, there
is a nontrivial convergence radius. When $p \geq q+2$, the series
converges everywhere except at $0$, with one notable exception, made 
by the polynomial hypergeometrics, i.e.
those for which some $a_i$ is a negative integer, which forces the series
to terminate after a finite number of terms.

This definition of a hypergeometric function assumes an argument $(x_1, \ldots, x_m) \in \mathbb{R}^m$; similarly one can extend the definition to hypergeometric functions of arguments in $(x_1, \ldots, x_m; y_1, \ldots, y_m; \ldots) \in  \mathbb{R}^m \times \mathbb{R}^m \times ...$ by inserting an additional \\ $C_{\kappa}^{\alpha}(y_1, \ldots y_m)/C_{\kappa}^{\alpha}(1, \ldots, 1)$ for each extra vector in $\mathbb{R}^m$.

Hypergeometric function\index{hypergeometric functions!definition}s provide answers to many statistics and statistics-related questions; below are two examples.

\begin{enumerate} \item Krishnaiah\index{Krishnaiah, P.R.} and Chang\index{Chang, T.C.} \cite{krishnaiah_chang} have proved in 1971 that the density of the smallest root of a real $(\alpha = 2)$ Wishart matrix with $m$ variables and $n$ degrees of freedom such that $p = \frac{n-m-1}{2}$ is an integer is proportional to
\[
\rho(x) = x^{pm} ~e^{-xm/2} ~{}_2F_{0}(-p, \frac{m+2}{2}; -2I_{m-1}/x)~.
\]

Note that the joint eigenvalue density of the matrix described above is given by $d\mu_{L}^{\alpha}$ with $\alpha = 2$ and $\gamma = p$.

In \cite{dumitriu03th} we extend this to any $\alpha$ and any positive integer $\gamma = p$. We obtain that for this case the density of the smallest eigenvalue is proportional to 
\begin{eqnarray} \label{seld}
\rho(x) = x^{pm} ~e^{-xm/2} ~{}_2F_{0}^{\alpha}(-p, \frac{m}{\alpha}+1; -2I_{m-1}/x)~.
\end{eqnarray}

\item 
The largest eigenvalue ($l_1$) distribution for a Wishart real matrix with 
$m$ variables and $n$ degrees of freedom ($\alpha=2$,  $\gamma = 
\frac{n-m-1}{2}$) can be expressed as \[
P[l_1<x] = \frac{\Gamma_m\big[\frac{1}{2}(m+1)\big]}{\Gamma_m\big[\frac{1}{2}(n+m+1)\big]} \left(\frac{x}{2}\right)^{mn/2} {}_1F_{1}\big( \frac{1}{2}n, \frac{1}{2}(n+m+1); -\frac{1}{2}xI_m\big)~.
\]

The above is a corollary of a stronger theorem proved by Constantine\index{Constantine, A.G.} \cite{constantine_noncentral}, and it can also be found in Muirhead\index{Muirhead, R.} \cite[page 421]{muirhead82a}. 

This result generalizes to any $\alpha$ and $\gamma$ (as noted in \cite{dumitriu03th}) to
\begin{eqnarray*}
P[l_1<x] & = & \frac{\Gamma_m\big[\frac{1}{\alpha}(m-1)+1\big]}{\Gamma_m\big[\gamma+\frac{2}{\alpha}(m-1)+2\big]} \left(\frac{x}{2}\right)^{m(\gamma+(m-1)/\alpha+1)} ~~\times \\
& & ~~~~\times~ {}_1F_{1}\big( \gamma+\frac{(m-1)}{\alpha}+1, ~a+\frac{2}{\alpha}(m-1)+2; ~-\frac{1}{2}xI_m\big)~
\end{eqnarray*}

%Hertz (1955) and Constatine (1963) have independently proved that for a 
non-central real ($\alpha = 1$) Wishart matrix $A = Z'Z$, with $Z$ a 
matrix of independent Gaussians with mean $M$ and variance $I_n \times 
\Sigma$, and with matrix of noncentrality parameters $\Omega = 
\Sigma^{-1}M^{T}M$, the moments of the determinant\index{moments of the determinant}
%\[
%E \Big[ (\det A)^r \Big] = (\det \Sigma)^r ~ 2^{mr}~ \frac{\Gamma_m \big(\frac{1}{2}n+r \big)}{\Gamma_m  \big(\frac{1}{2}n \big)} ~{}_1F_{1}(-r; \frac{1}{2}n; -\frac{1}{2} \Omega)~.
%\] 
\end{enumerate}

\section{Software}

\subsection{The model for MOPS}

We have initially chosen as a model for \textbf{MOPS} the Symmetric Functions (\textbf{SF}) package by John Stembridge\footnote{The \textbf{SF} package can be found at the URL \hspace{.25cm} http://www.math.lsa.umich.edu/$\sim$jrs/maple.html\#SF.}.  Our library is compatible with \textbf{SF}, and our procedures \textbf{m2p}, \textbf{p2m}, and \textbf{m2m} are designed to complement the \textbf{SF} procedures \textbf{tom} and \textbf{top} (for a comparison, see Section \ref{perf}). Though in time our vision of \textbf{MOPS} has changed, we are grateful to John Stembridge for his ground-breaking work.

\subsection{System requirements and installation guide} 

Our library was initially developed for Maple 7, and later for Maple 8. Experience has shown that it is also 
compatible with Maple 9.

We have developed \textbf{MOPS} on various Unix machines and one Windows
XP machine; the same version of the library is compatible with both
operating systems. Below we provide an installation guide, which can also 
be found on the MOPS webpage, located at 
http://www.math.berkeley.edu/$\sim$dumitriu/mopspage.html .

\vspace{.5cm}

\emph{For Unix users}: download the file \textbf{mops.zip} into your home directory. Unzip the file using the command
\[
\mbox{unzip MOPS.zip}
\]

This should create a new MOPS directory in your home directory; the new MOPS directory contains a subdirectory named \textbf{Help\ Files} and 4 files, named maple.hdb, MOPS.ind, MOPS.lib, and MOPS.mws. 
The last file contains an archive of all the procedures we wrote for \textbf{MOPS}. 

Return to your home directory (e.g. `/home/usr/vis/joesmith/`), and create a .mapleinit file; write in 
\begin{eqnarray*}
\mbox{new$\_$libname} &:=& \mbox{`/home/usr/vis/joesmith/MOPS`}~; \\
\mbox{libname} & :=& \mbox{libname}, ~\mbox{new$\_$\mbox{libname}}~;
\end{eqnarray*}
and then save the .mapleinit file. 

All that is left is to call the library (in Maple) using the standard Maple command for libraries, i.e. 
\begin{eqnarray*}        
> \mbox{with(MOPS)}~;  
\end{eqnarray*}
each time you need to use the library. 

\vspace{.5cm}

\emph{For Windows users}: users:  place the downloaded file in your C:$\backslash$ directory (or in a more appropriate place of your choosing).

Unzip the file using Winzip; this should create a new C:$\backslash$MOPS directory in your home directory; the MOPS directory contains a subdirectory entitled Help\ Files and 4 files, named maple.hdb, MOPS.ind, MOPS.lib, and MOPS.mws . The last file contains an archive of all the procedures we wrote for MOPs. 

\begin{itemize} \item[1.] In the $~\backslash$Maple$\backslash$bin folder, create a .mapleinit file. In it you should write
\begin{eqnarray*}
\mbox{new$\_$libname} &:=& \mbox{`C:$\backslash \backslash$MOPS`}~; \\
\mbox{libname} & :=& \mbox{libname}, ~\mbox{new$\_$\mbox{libname}}~;
\end{eqnarray*}
and then save the file.

You will need to call (in Maple) the library each time you need to use it, using the standard command
\begin{eqnarray*}        
> \mbox{with(MOPS)}~;  
\end{eqnarray*}

\item[2.] For some unknown reasons, the instructions in variant 1 do not always work on a Windows XP machine. In that case, you will have to type in the pathway for the library, each time you will need to use it. Upon opening a Maple window, you will have to type in 
\begin{eqnarray*}
& >& \mbox{new$\_$libname} := \mbox{`C:$\backslash \backslash$MOPS`}~; \\
& >& \mbox{libname} := \mbox{libname}, ~\mbox{new$\_$\mbox{libname}}~; \\
& >& \mbox{with(MOPS)};
\end{eqnarray*}
each time you need the library.
\end{itemize}

\vspace{.5cm}

\emph{For Mac users}: the instructions are similar to the instructions for Windows.

\vspace{1cm}

Regardless of the Operating System, we suggest you perform a check before you start using the library. For example, if you type in
\begin{eqnarray*}
&>& \mbox{jack(a, }[3],~2, ~'\mbox{P}')~;
\end{eqnarray*}
the answer should be
\begin{eqnarray*}
\mbox{m}[3]+ \frac{3 ~\mbox{m}[2,1]}{1+2\mbox{a}}
\end{eqnarray*}

\subsection{Routines: name, syntax, and description}

We give here a list of the 30 routines, and a brief description of what they do; for more extensive mathematical explanation we refer to Section \ref{math}. Note that most of them are set up to do calculations both symbolically and numerically.

We use the following notations: \begin{enumerate} 
\item $\kappa, ~\sigma$  for partitions, 
\item $k, ~s$ for the corresponding partition sums, 
\item $n, ~m$ for the number of variables, 
\item $\alpha$ for the Jack parameter, 
\item $i, j$ for the location of a square in a partition, 
\item $A_p,~B_q$ for two lists of real parameters, one of length $p$ and the other of length $q$,
\item $l$ for a computational limit, 
\item $\gamma, g_1, g_2$ for additional Laguerre or Jacobi parameters,
%\item $x$ for a list of variables, 
\item $[x]$ for either a number or a list of variables,
\item $r$ for a real parameter,
\item \textit{N} for the normalization, 
\item \textit{exp} for an expression.
\end{enumerate}

Some parameters are optional.

Three of the routines, namely, \textbf{m2p}, \textbf{p2m}, and \textbf{m2m}, are alternatives to the routines \textbf{tom} and \textbf{top} from the \textbf{SF} package to the conversion between the monomial and the power sum basis. For a comparison between our routines and their \textbf{SF} counterparts, see Section \ref{perf}. %Our versions are almost always faster, because the algorithms are targeted (not as general as the ones used in \textbf{SF}).

\begin{figure}[ht]
\caption{Dependence graph for the procedures of MOPS.} \label{dependence}
\begin{center}
\epsfig{figure = 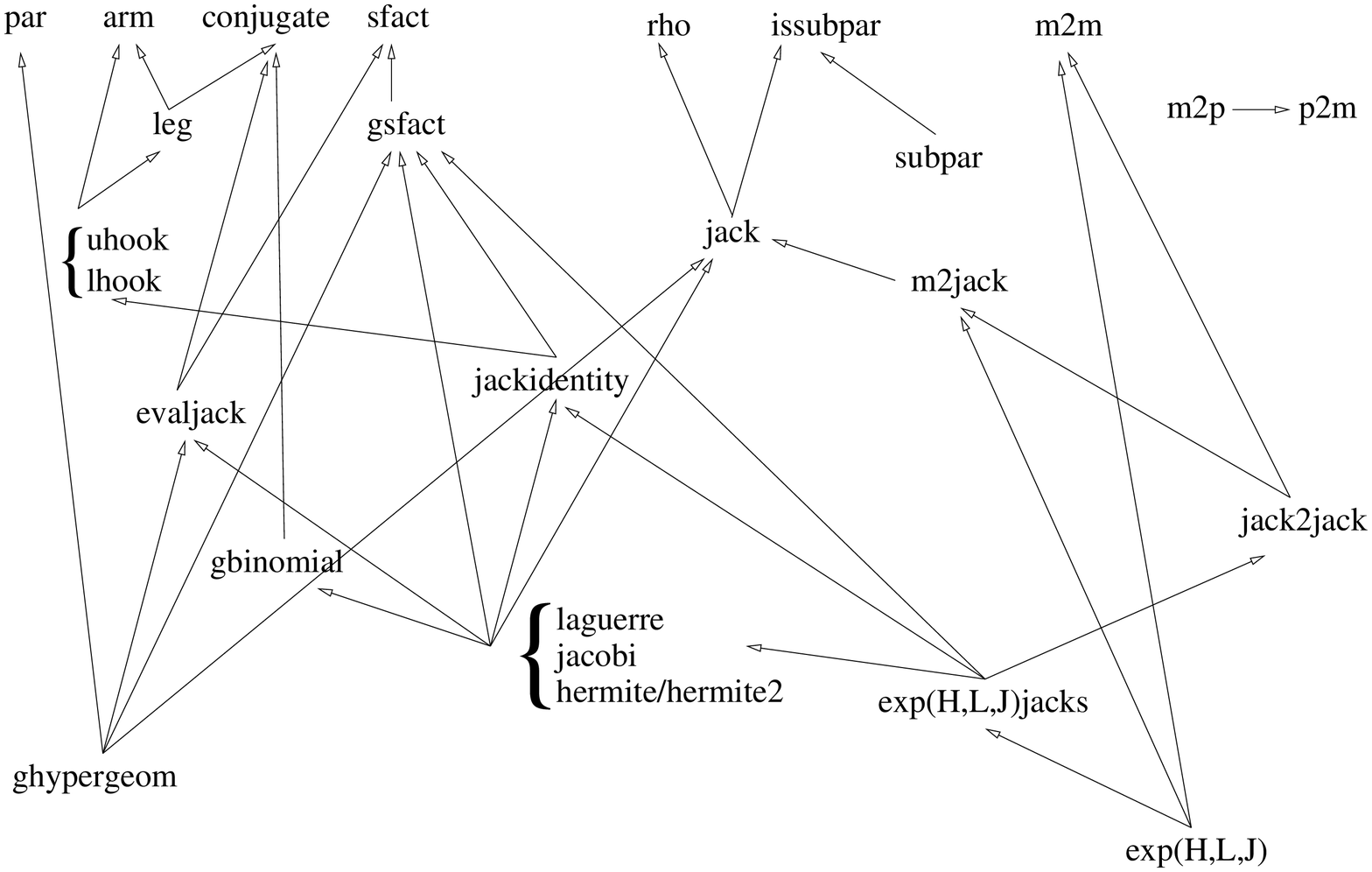, width = 16cm}
\end{center}
\end{figure}

Some of the routines are used as building blocks for others; for a tree description of the dependencies, see Figure \ref{dependence}.

%The procedures \textbf{arm}, \textbf{conjugate}, \textbf{dominate}, \textbf{issubpar}, \textbf{leg}, \textbf{lhook}, \textbf{par}, \textbf{subpar}, and \textbf{uhook} are building blocks, and have similar counterparts in the \textbf{SF} package.

\vspace{1cm}

\hspace{-1cm}\begin{tabular}{lll}
\textit{Procedure} & \textbf{Syntax} & \textit{Description} \\ \hline
\textbf{arm} & arm$(\kappa, i,j)$ & the arm-length of a partition at a square \\ \hline
\textbf{conjugate} & conjugate$(\kappa)$ & the conjugate partition \\ \hline
\textbf{expH} & expH$(\alpha, exp, n)$ & the expected value of an expression in terms of  \\
& & monomials with respect to the $2/\alpha$-Hermite \\ 
& & distribution with $n$ variables\\ \hline
\textbf{expHjacks} & expHjacks$(\alpha, exp, n)$ & the expected value of an expression in terms of Jack \\
& & polynomials with respect to the $2/\alpha$-Hermite \\
& & distribution with $n$ variables\\ \hline
\textbf{expJ} & expJ$(\alpha, exp, g_1, g_2, n)$ & the expected value of an expression \\
& & in terms of monomials with respect to the \\
& & $2/\alpha, ~g_1, ~g_2$-Jacobi distribution with $n$ variables \\ \hline

\textbf{expJjacks} & expJjacks$(\alpha, exp, g_1, g_2, n)$ & the expected value of an expression \\
& & in terms of Jack polynomials with respect to the \\
& & $2/\alpha, ~g_1,~g_2$-Jacobi distribution with $n$ variables \\ \hline

\textbf{expL} & expL$(\alpha, exp, \gamma, n)$ & the expected value of an expression \\
& & in terms of monomials with respect to the \\
& & $2/\alpha,~\gamma$-Laguerre distribution with $n$ variables \\ \hline

\textbf{expLjacks} & expLjacks$(\alpha, exp, \gamma, n)$ & the expected value of an expression \\
& & in terms of Jack polynomials with respect to the \\
& & $2/\alpha, ~\gamma$-Laguerre distribution on $n$ variables \\ \hline

\textbf{gbinomial} & gbinomial$(\alpha, \kappa, \sigma)$ & the generalized binomial coefficient \\ \hline

\textbf{ghypergeom} & ghypergeom$(\alpha, A_p, B_q, [x], '\!N', l)$ & the generalized hypergeometric function \\ 
& & corresponding to parameter lists $A_p, B_q$, evaluated \\
& & at the point $x \in \mathbb{R}^n$ (or written \\
& & symbolically for $n$ variables), or (optional) \\
& & as a series ``cut'' after terms that sum up to $l$ \\  \hline

\textbf{gsfact} & gsfact$(\alpha, r, \kappa)$ & the generalized shifted factorial (or generalized \\
& & Pochhammer symbol) \\ \hline

\textbf{hermite} & hermite$(\alpha, \kappa, [x],'\!N')$ & the multivariate Hermite polynomial \\
& & written in Jack polynomial basis\\ \hline

\textbf{hermite2} & hermite2$(\alpha, \kappa, [x], '\!N')$ & alternative way of computing the multivariate \\
& & Hermite polynomial written in Jack polynomial basis\\ \hline

\textbf{issubpar} & issubpar$(\sigma, \kappa)$ & checks if $\sigma$ is a subpartition of $\kappa$\\ \hline
%\end{tabular}

%\begin{tabular}{lccl}
%\textit{Procedure} && & \textit{Description} \\ \hline
\textbf{jack} & jack$(\alpha, \kappa, [x], 'N')$ & the Jack polynomial as a linear combination \\
& & of monomials\\ \hline
\textbf{jack2jack} & jack2jack$(\alpha, exp, n)$ & converts a polynomial expression involving Jack \\
& & polynomials into a linear combination of \\
& & Jack polynomials \\ \hline
\textbf{jackidentity} & jackidentity$(\alpha, \kappa, m)$ & the Jack polynomial evaluated at $x_i = 1$, $i=1..m$ \\ \hline
\textbf{jacobi} & jacobi$(\alpha, \kappa, g_1, g_2, [x], '\!N')$ & the multivariate Jacobi polynomial as a linear \\
& & combination of Jack polynomials\\ \hline

\textbf{laguerre} & laguerre$(\alpha, \kappa, g, [x], '\!N')$ & the multivariate Laguerre polynomial as a linear \\
& & combination of Jack polynomials \\ \hline
\textbf{leg} & leg$(\kappa, i,j)$ & the leg-length of a partition at the square $(i,j)$ \\ \hline
\textbf{lhook} & lhook$(\alpha, \kappa)$ & the lower hook of a partition \\ \hline

\end{tabular}

\hspace{-1cm} \begin{tabular}{lll} 
\textit{Procedure} & \textbf{Syntax} & \textit{Description} \\ \hline

\textbf{m2jack} & m2jack$(\alpha, exp, n)$ & converts an expression involving monomials in \\ 
& & Jack polynomial basis \\ \hline
\textbf{m2m} & m2m$(exp, n)$ & converts an expression involving monomials \\
& & to a linear combination of monomials \\ \hline
\textbf{m2p} & m2p$(exp)$ & converts an expression involving monomials\\
& & to a linear combination of power sum functions \\ \hline
%\end{tabular}

%\hspace{-1cm} \begin{tabular}{lll} 
%\textit{Procedure} & \textbf{Syntax} & \textit{Description} \\ \hline

\textbf{p2m} &  p2m$(exp, n)$ & converts an expression involving power sum functions \\
&  & to a linear combination of monomials\\ \hline
\textbf{par} &  par$(k)$ & produces and lists all partitions of a given integer \\ \hline
\textbf{rho} & rho$(\alpha, \kappa)$ & the $\rho$ function of a partition \\ \hline
\textbf{sfact} & sfact$(r, k)$ & the shifted factorial (Pochhammer symbol) \\ \hline
\textbf{subpar} & subpar$(\kappa)$ & produces and lists all subpartitions of a given partition \\ \hline
\textbf{uhook} & uhook$(\alpha, \kappa)$& the upper hook of a partition  \\ \hline
\end{tabular}

\subsection{Computing Expected Values / Evaluating Integrals} \label{compexp}

Let $P_{\lambda}(x_1, \ldots, x_m)$ be a symmetric polynomial in $m$ variables, with highest order term corresponding to the partition $\lambda$. To compute the expected value of $P_{\lambda}$ with respect to one of the distributions $d\mu_{J}^{\alpha}$ (\ref{jac_density}), $d\mu_{L}^{\alpha}$ (\ref{lag_density}), or $d\mu_{H}^{\alpha}$ (\ref{herm_density}), we write 
\[
P_{\lambda}(x_1, \ldots, x_m) = \sum_{\kappa} c_{\kappa, \alpha} C_{\kappa}^{\alpha} (x_1, \ldots, x_m)~,
\]
and by applying the linearity of expectation, we obtain 
\[
E \big[P_{\lambda}(x_1, \ldots, x_m) \big]= \sum_{\kappa} c_{\kappa, \alpha} E \big[C_{\kappa}^{\alpha} (x_1, \ldots, x_m) \big]~.
\]

In the univariate case the Jack polynomials are simply monomials, and we have the following (well-known) moments for the Hermite, Laguerre, and Jacobi weight functions:
\begin{eqnarray*}
\!\!\!\frac{1}{\sqrt{2\pi}}\int_{\mathbb{R}} x^k e^{-x^2/2} dx  & = & (2k-1)!!  =  (-1)^{k/2} H_{k}(0)~, \\
\!\!\!\frac{1}{\Gamma(\gamma+1)}\int_{[0, \infty)} x^k x^{\gamma} e^{-x} dx  & = & (\gamma+1)_k  = L_{k}^{\gamma}(0)~, ~~~~\mbox{and}\\
\!\!\!\frac{\Gamma(2+a+b)}{\Gamma(a+1)\Gamma(b+1)}\int_{[0,1]} x^k x^a (1-x)^b dx  & = &\frac{(a+1)_k\Gamma(a+b+2)}{\Gamma(a+1) \Gamma(a+b+k+2)} = J_{k}^{a,b}(0)~.
\end{eqnarray*}
In the above, $k \geq 0$.

A similar triad of formulas is can be established for the multivariate case. In the Laguerre and Jacobi cases, the univariate formulas generalize easily:
\begin{eqnarray} \label{expL}
\!\!\!\!\!\!\!\!\!\!\!\!\int_{[0, \infty)^m} \!\!\!\!C_{\kappa}^{\alpha} (x_1, \ldots, x_m) d\mu_{L}^{\alpha}(x_1, \ldots, x_m) &\!\!\! = \!\!\!& (\gamma+\frac{m-1}{\alpha}+1)_{\kappa} C_{\kappa}^{\alpha}(I_m) = L_{\kappa}^{\alpha, \gamma}(0)~,\\
\!\!\!\!\!\!\!\!\!\!\!\!\label{expJ}\int_{[0,1]^m} \!\!\!\!C_{\kappa}^{\alpha} (x_1, \ldots, x_m)  d\mu_{J}^{\alpha}(x_1, \ldots, x_m)  & \!\!\!= \!\!\!& \frac{(g_1\!+\!\frac{m-1}{\alpha}\!+\!1)_{\kappa}}{(g_1\!+\!g_2\!+\!\frac{2}{\alpha}(m-1)\!+\!2)_{\kappa}} C_{\kappa}^{\alpha}(I_m) = J_{\kappa}^{\alpha, g1, g2}(0).~~~~
\end{eqnarray}

For a good reference for the first formula, see 
Forrester\index{Forrester, P.J.} and Baker\index{Baker, T.} \cite{Forrester_poly}; the second one was obtained by Kadell\index{Kadell, K.} \cite{kadell_jacks}.

For the Hermite case,
\begin{eqnarray} \label{expH} 
\int_{\mathbb{R}^n} C_{\kappa}^{\alpha} (x_1, \ldots, x_m) d\mu_{H}^{\alpha} = (-1)^{k/2}  H_{\kappa}^{\alpha}(0)~,
\end{eqnarray}
but to the best of our knowledge, no simpler closed-form formula is known. We compute the right hand side as the 0th order coefficient of the polynomial $H_{\kappa}^{\alpha}(x_1, \ldots, x_m)$, using formula \eqref{mine}. Note that if $\kappa$ sums to an odd integer, the above is trivially $0$.

The procedures \textbf{expHjacks}, \textbf{expLjacks}, and \textbf{expJjacks} compute the expected value of an expression (allowing not only for addition, but also for multiplication and powers) involving Jack polynomials (all having the same parameter $\alpha$ as the distribution). They consist of the two steps described in the first paragraph of this section: the first one is reducing the expression to a weighted sum of Jack polynomials (if the expression \emph{is} a weighted sum of Jack polynomials, this step is skipped), and the second step is replacing each Jack polynomial with its expected value, using the formulas \eqref{expH}, \eqref{expL}, and \eqref{expJ}. 

The procedures \textbf{expH}, \textbf{expL}, and \textbf{expJ} compute the expected value of an expression involving monomials (allowing for addition, multiplication, and powers), and there are three steps involved: the first is to write  the expression in monomial basis, the second -- to rewrite the result in Jack polynomial basis, and the third is to  replace the Jack polynomials by their expectations, using  \eqref{expH}, \eqref{expL}, and \eqref{expJ}.

\vspace{.25cm}

\noindent \textbf{Example.} Suppose we want to compute the expected value of 
\[ z(\alpha, x_1, x_2, x_3) := J_{[2,1]}^{\alpha}(x_1, x_2, x_3) 
C_{[1,1,1]}^{\alpha}(x_1, x_2, x_3)\] over the $2/\alpha$-Hermite distribution. First we have to express $z$ as a linear combination of Jack ``C'' Polynomials. Note that the number of variables, as well as $\alpha$, must be the same in the two terms of $z$.

First, we express the two terms in monomial basis:
\begin{eqnarray*}
J_{[2,1]}^{\alpha}(x_1, x_2, x_3) & = & (2+\alpha)~m_{[2,1]}(x_1, x_2, x_3) + 6~m_{[1,1,1]}(x_1, x_2, x_3)~, \\
C_{[1,1,1]}^{\alpha}(x_1, x_2, x_3) & = & \frac{6\alpha^2}{(1+\alpha)(2+\alpha)}~m_{[1,1,1]}(x_1, x_2, x_3).
\end{eqnarray*}

Their product thus becomes a linear combination of sums of products of two monomials, which are in turn converted in linear combinations of monomials. Note that here we use the fact that there are three variables:
\begin{eqnarray*}
m_{[2,1]}(x_1, x_2, x_3) ~m_{[1,1,1]}(x_1, x_2, x_3) & = & m_{[3,2,1]}(x_1, x_2, x_3)~, ~~~\mbox{while} \\
m_{[1,1,1]}(x_1, x_2, x_3)^2 & = & m_{[2,2,2]}(x_1, x_2, x_3)~.
\end{eqnarray*}

Putting it all together, in monomial basis,
\begin{eqnarray*}
z(\alpha, x_1, x_2, x_3) &=& \frac{6\alpha^2}{1+\alpha}~m_{[3,2,1]}(x_1, x_2, x_3) ~~+ \\
& & ~~+~~ \frac{36\alpha^2}{(1+\alpha)(2+\alpha)}~m_{[2,2,2]}(x_1, x_2, x_3)~.
\end{eqnarray*}

All that is left now is to convert from monomial basis back to Jack polynomial basis. We obtain that 
\begin{eqnarray*}
z(\alpha, x_1, x_2, x_3) &=& \frac{1}{120} \frac{(2+3\alpha)(1+2\alpha)^2}{\alpha(1+\alpha)}~C_{[3,2,1]}^{\alpha}(x_1, x_2, x_3) 
\end{eqnarray*}

We are now able to finish the work: 
\begin{eqnarray*}
E_{H} \big[z(\alpha, x_1, x_2, x_3) \big] = - \frac{36(\alpha-1)(\alpha+3)}{(1+\alpha)(2+\alpha)}~.
\end{eqnarray*}

\section{Complexity bounds and running times} \label{perf}

In this section we will analyze the performance of the main algorithms, which we divide into four parts: \begin{enumerate} \item algorithms that compute polynomials; \item algorithms that evaluate integrals; \item conversion algorithms; \item numerical algorithms. \end{enumerate}

Our complexity bounds are upper bounds, but we believe many of them to be asymptotically correct. They work well for the numerical evaluation of the parameters involved (i.e. $\alpha$, $m$, $\gamma$, $g_1$, $g_2$); symbolic evaluation of the polynomials is considerably slower.We are not aware of the existence of a good symbolic performance model for Maple, and hence it would be difficult to predict how much slower symbolic evaluation is than numerical evaluation. Once parameters are introduced (like $m$, the number of variables, or $\alpha$, the Jack parameter), the quantities to be computed become rational functions of these parameters, of degrees that can go up to the partition size $|\kappa|$. Storage then becomes an issue, hence one would expect that the running times for symbolic evaluation would be orders of magnitude slower than for numerical evaluation, since the coefficients we deal with must be written and stored on ``slow'' memory (e.g. disk space), and the ``transport'' time to and from ``slow'' memory greatly increases the overall running time. 

For each algorithm we provide a complexity analysis, and we illustrate the performance in practice by providing running times for different tests (both numerical and symbolic); then we examine the running times and draw a set of conclusions.

Each time we use $N/A$ for an entry in a running times table, we have done so because that particular computation has exhausted the memory available to Maple, and hence (regardless of the time it took up to that point) the computation was not finished.

The computer on which we have performed our tests is a Pentium 4 by Dell, 1.8 Ghz, 512 MB; the version of Maple used for the tests is Maple 8.

The last thing worth mentioning is that Maple has an \emph{option remember}, that is it allows for storage and recall of a quantity that was computed previously, and that MOPS is taking advantage of that.

\subsection{Algorithms that compute polynomials}

In this category we have the algorithms that evaluate \textbf{jack}, \textbf{gbinomial}, \textbf{hermite}, \textbf{laguerre}, and \textbf{jacobi}. We analyze here \textbf{gbinomial}, though it is not a polynomial in $(x_1, \ldots, x_m)$, because it is the main building block for \textbf{hermite}, \textbf{laguerre}, and \textbf{jacobi}, and its complexity determines their computational complexity. 

Throughout this section, we will follow the notations given in Table \ref{sapte}.

\begin{table}[ht]
\begin{center} 
\begin{tabular}{|c||l|}
\hline
$k = |\kappa|$ & \hspace{.1cm} size of partition $\kappa$ \\
$s = |\sigma|$ & \hspace{.1cm} size of partition $\sigma$ \\
$l = $length$(\kappa)$ & \hspace{.1cm} length of partition $\kappa$ \\
$n$   & \hspace{.1cm} number of variables used for computation \\
\hline
$D_{\kappa}$ & \hspace{.1cm} number of partitions of $k$ smaller in lexicographical ordering than $\kappa$ \\
$P_{\kappa}$ & \hspace{.1cm} number of partitions of $k$ dominated by $\kappa$ \\
$P_{[k]}$ &  \hspace{.1cm} number of partitions of the number $k$ (each partition of \\
& \hspace{.5cm} $k$ is dominated by $[k]$) \\
$U_{\kappa}$ & \hspace{.1cm} number of subpartitions of $\kappa$ \\
$U_{\kappa, \sigma}$ & \hspace{.1cm} number of subpartitions of $\kappa$ which are superpartitions\\
& \hspace{.5cm} for $\sigma$ (this implies $\sigma$ is a subpartition of $\kappa$)\\
$A_{\kappa}$ & \hspace{.1cm} number of subpartitions of $\kappa$ which sum to a number with \\
& \hspace{.5cm} the same parity with $k$ \\
\hline
\end{tabular} 
\end{center}
\caption{Notations to be used throughout Section \ref{perf}.} \label{sapte}
\end{table}

To make estimates, we have used Ramanujan's formula:
\begin{eqnarray} \label{Ramanujan} 
P_{[k]} \sim \frac{1}{4k \sqrt{3}} ~ e^{\pi \sqrt{2k/3}}~~, 
\end{eqnarray}
and the inequalities 
\begin{eqnarray} \label{ine}
A_{\kappa} \llq U_{\kappa} \llq P_{\kappa} \llq P_{[k]} ~~~\mbox{and}~~~ U_{\kappa, \sigma} \llq P_{[k]}
\end{eqnarray}
for asymptotical estimates.

\begin{enumerate}
\item \textit{\textbf{jack}}. 
The algorithm uses recurrence \eqref{recur_jack}, together with the `boundary conditions' $c_{\kappa, \lambda} = 0$ if $\kappa \not \succeq \lambda$ in dominance ordering, and $c_{\kappa, \kappa} = \frac{\alpha^{k} k!}{c'(\kappa, \alpha)}$. The length of the recurrence is at most $O(k_1 {k+1 \choose 2})$, with $k_1$ being the first entry in the partition, and the algorithm will check each of the possible partitions $\mu$ (at most $k_1 {k+1 \choose 2}$) to see if they are dominated by $\kappa$ and dominating $\lambda$ (this involves $l$ additions and $l$ comparisons). The rest of the computation has complexity $O(k)$.
\index{Jack polynomials!algorithm}

Thus the complexity of the algorithm is $O(k_1 k^3 P_{\kappa})$. 

Using the inequalities \eqref{ine}, the best asymptotical upper bound we can get is for the complexity of computing a Jack polynomial is thus $O(k^3 e^{\pi \sqrt{2k/3}})$, which is super-polynomial.

%The algorithm uses the recurrence (\ref{recur_jack}), together with the ``boundary'' conditions listed there. Let $l$ be the length of the partition $\kappa = (k_1, k_2, \ldots, k_l)$, and let $k = |\kappa|$. The length of recurrence (\ref{recur_jack}) is at most $k_1 {k \choose 2}$, and since there are exactly $D_{\kappa}$  coefficients to compute, it follows that the total work done to compute \textbf{jack(a, kappa)} is $O(k_1 k^2 D_{\kappa})$. Using that $k_1 < k$ and $D_{\kappa} \leq P_{k}$, we obtain by using Ramanujan's formula $P_k \sim \frac{e^{\pi \sqrt{2/3} \sqrt{k}}}{4k \sqrt{3}}$, that the running time is bounded from above by $O(k^2 e^{\pi \sqrt{2/3} \sqrt{k}})$.

%This is an upper bound, but it is within a constant of optimal when $\kappa$ has a small number of parts; in particular when $\kappa = [k-1,1]$ -- not when $\kappa = [k]$, because there is a closed-form formula for $J_{\kappa}^{\alpha}$ which allows for only $O( e^{\pi \sqrt{2/3} \sqrt{k}})$ work.

Below we illustrate the running times for both numerical and symbolic computations.
 For numerical computations, we have chosen to make $\alpha = 1$, so that the Jack polynomials\index{Jack polynomials!algorithm} are the Schur functions. Note that we do not test the partition $[k]$; for that particular partition we have a closed-form formula for the Jack polynomial, due to Stanley \cite{stanley_jacks}, which has complexity  $O(kP_k) \sim O(e^{\pi \sqrt{2k/3}})$.

%\begin{tabular}{|l|l|r|r|}
%\hline
%Partition sum & Partition &  \hspace{.25cm} Runtime    & \hspace{.25cm} Runtime \\
%              &           & (in seconds) & (in seconds) \\
%              &           & $\alpha = 2~~~$ & $\alpha~~~~~$ \\ \hline
%$k=15$ & $\kappa = [14,1]$ & $4.63$ & $7.54$\\ 
%       & $\kappa = [8,7]$ & $3.31$ & $5.00$\\
%       & $\kappa = [3,3,3,3,3]$ & $0.77$ & $0.92$\\ \hline
%$k=20$ & $\kappa = [19,1]$ &$32.22$ & $48.42$\\ 
%       & $\kappa = [10,10]$ &$21.92$ & $32.36$\\
%       & $\kappa = [4,4,4,4,4]$ &$5.80$ &$6.89$ \\ \hline
%$k=25$ & $\kappa = [24,1]$ &$178.34$ &$284.49$ \\ 
%       & $\kappa = [8,8,8,1]$ & $76.07$& $112.24$\\
%       & $\kappa = [6,6,6,6,1]$ & $46.77$& $61.57$\\ \hline
%$k=30$ & $\kappa = [29,1]$ & $1052.99$& $1985.00$\\ 
%       & $\kappa = [10,10,10]$ & $380.71$& $628.85$\\
%       & $\kappa = [6,6,6,6,6]$ & $150.22$& $190.85$\\ \hline
%\end{tabular}

\begin{table}[ht!]
\begin{center}
\begin{tabular}{|l||l||c||c||c|}
\hline
$k$ & $\kappa$ & Running time, $\alpha = 1$ & Running time, $\alpha$ symbolic & Ratio \\ \hline
$15$ & $\kappa = [14,1]$ & $2.48$ & $4.54$ & $1.83$ \\ 
       & $\kappa = [8,7]$ & $1.79$ & $3.17$ & $1.77$ \\
       & $\kappa = [3,3,3,3,3]$ & $0.39$ & $0.50$ & $1.28$ \\ \hline
$20$ & $\kappa = [19,1]$ &$16.97$ & $30.45$ & $1.79$ \\ 
       & $\kappa = [10,10]$ &$11.53$ & $20.32$ & $1.76$ \\
       & $\kappa = [4,4,4,4,4]$ &$2.91$ &$4.02$ & $1.38$ \\ \hline
$25$ & $\kappa = [24,1]$ &$93.42$ &$189.66$ & $2.03$ \\ 
       & $\kappa = [9,8,8]$ & $46.85$& $79.85$ & $1.70$ \\
       & $\kappa = [5,5,5,5,5]$ & $16.08$& $24.18$ & $1.50$ \\ \hline
$30$ & $\kappa = [29,1]$ & $634.32$ & $1819.65$ & $2.86$ \\ 
       & $\kappa = [10,10,10]$ & $214.10$& $418.19$ & $1.95$ \\
       & $\kappa = [6,6,6,6,6]$ & $73.54$& $113.55$ & $$ $1.54$ \\ \hline
\hline
\end{tabular}
\end{center}
\caption{Running times (in seconds) for the Jack polynomial computation.} \label{table_jack_times}
\end{table}
%\begin{tabular}{|l|l|r|r|r|}
%\hline
%Partition sum & Partition &  \hspace{.25cm} Runtime    & \hspace{.25cm} Runtime & Ratio \\
%              &           & (in seconds) & (in seconds) & \\
%              &           & $\alpha = 2~~~$ & symbolic $\alpha$ & \\ \hline
%$k=15$ & $\kappa = [14,1]$ & $2.44$ & $4.57$ & $1.87$ \\ 
%       & $\kappa = [8,7]$ & $1.79$ & $3.09$ & $1.72$ \\
%       & $\kappa = [3,3,3,3,3]$ & $0.39$ & $0.52$ & $1.33$ \\ \hline
%$k=20$ & $\kappa = [19,1]$ &$17.80$ & $27.92$ & $1.56$ \\ 
%       & $\kappa = [10,10]$ &$11.79$ & $18.79$ & $1.59$ \\
%       & $\kappa = [4,4,4,4,4]$ &$3.04$ &$3.63$ & $1.19$ \\ \hline
%$k=25$ & $\kappa = [24,1]$ &$101.34$ &$170.97$ & $1.68$ \\ 
%       & $\kappa = [8,8,8,1]$ & $41.41$& $65.17$ & $1.57$ \\
%       & $\kappa = [6,6,6,6,1]$ & $25.60$& $35.09$ & $1.37$ \\ \hline
%$k=30$ & $\kappa = [29,1]$ & $752.96$& $1656.98$ & $2.20$ \\ 
%       & $\kappa = [10,10,10]$ & $237.10$& $507.19$ & $2.13$ \\
%       & $\kappa = [6,6,6,6,6]$ & $82.24$& $110.38$ & $$ $1.34$ \\ \hline
%\end{tabular}

\vspace{.25cm}
 
\begin{remark}
Note that the ratio of the running times increases when the partition size increases. At $k=30$, the number of partitions is $5604$, and each of the monomial coefficients is a rational function of $\alpha$. Issues like storage and memory access become important, and influence negatively the running times. Another important factor is that in order to make things easier to store and access, not to mention easier to read and interpret, we use the procedures ``simplify'' and ``factor'', which are relatively costly.
\end{remark}

\vspace{.25cm}

\textit{Extrapolation}. Since the speed/memory of a top-of-the-line computer seems to go up by a factor of $10^3$ every $10$ years, one can predict that within a decade, using MOPS, computing $J_{(59,1)}^{\alpha}$ will take about $30$ minutes.

\item \textit{\textbf{gbinomial}}. We use \eqref{recur_gbinom}, together with the boundary conditions listed in Section \ref{gbc} and with the contiguous binomial coefficient formula \eqref{cont_binom}. From \eqref{recur_gbinom}, it follows that computing a single contiguous binomial coefficient has complexity $O(k)$, and one needs to compute no more than $l$ such coefficients per subpartition $\tilde{\sigma}$ of $\kappa$ which is a superpartition of $\sigma$. 

Thus one immediately obtains the bound $O(kl U_{\kappa, \sigma})$ for the complexity of computing ${\kappa \choose \sigma}$. This is  smaller than $O(k^2 U_{\kappa, ~[1^2]})$.

Note that by computing ${\kappa \choose \sigma}$, one also obtains ${\kappa \choose \mu}$, for each $\sigma \subseteq \mu \subset \kappa$. So we have chosen for our tests to compute ${\kappa \choose [1,1]}$ for different $\kappa$, as this yields all the binomial coefficients having $\kappa$ as top partition (except ${\kappa \choose 2}$, but that requires only an additional complexity $O(kl)$).

By using the inequalities \eqref{ine}, we obtain an asymptotical upper bound of $O(k e^{\pi \sqrt{2k/3}})$ for computing \textit{all} the generalized binomial coefficients corresponding to partitions of $k$.

\begin{table}[ht!]
\begin{center}
\begin{tabular}{|l||l||c||c||c|}
\hline
$k$ & $\kappa$ & Running time, & Running time, & $U_{\kappa, ~[1^2]}$\\ 
    &          & $\alpha = 1$ & $\alpha$ symbolic & \\ \hline
$15$ &  $[6,4,2,2,1]$ & $0.22$ & $1.12$  &  $139$ \\
& $[3,3,3,3,3]$ & $0.05$ & $0.18$ & $56$ \\
& $[10,5]$ & $0.03$ & $0.15$ & $51$ \\
\hline
$20$ & $[6, 4, 3,2, 2,1,1,1]$ & $1.01$ &$6.68$  & $418$ \\
     & $[4,4,4,4,4]$ & $0.17$ & $0.6$ & $126$ \\
     & $[12,8]$  & $0.07$ & $0.28$ & $81$\\
\hline
$25$ & $[7,5,4,3,2,2,1,1]$ & $3.41$ & $23.37$ & $1077$ \\
     & $[5,5,5,5,5]$ &$0.41$ &$1.67$ & $252$\\
     & $[16,9]$ & $0.15$ & $0.62$ & $125$ \\
\hline
$30$ & $[8,6,4,3,2,2,1,1,1,1,1]$ & $11.87$& $89.61$ & $2619$ \\
     & $[6,6,6,6,6]$ & $0.91$ & $3.95$ & $462$ \\
     & $[20,10]$ & $0.24$ & $1.20$ & $176$ \\
\hline
\end{tabular}
\end{center}
\caption{Running times (in seconds) for the generalized binomial coefficient computation.} \label{table_gbinom_times}
\end{table}

\vspace{.25cm}

\begin{remark} Once again, size and length of the partition increase the symbolic running times; however, note that the running times are relatively small, even for partitions of $30$. We believe that the generalized binomial coefficients are rational functions of $\alpha$ which can always be factored in small-degree factors, so that they are easy to store and operate with.
\end{remark}

\item \textit{\textbf{jacobi}}.

To compute the Jacobi polynomials, we use the format of Section \ref{format_jacobi} and recurrence \eqref{recurrence_jacobi}. One can easily see that at each step, one needs to compute at most $l$ contiguous binomial coefficients, each of which has complexity $O(k)$; in addition, one needs to compute another at most $l$ binomial coefficients; each of these takes only $O(l)$, as the contiguous coefficients needed \emph{have already been computed} at the previous step. Thus the total complexity is $O(kl)$ (since $l \llq k$) at each step, for a total of $O(kl U_{\kappa,~[1^2]})$.

Hence computing numerically the Jacobi polynomials is comparable to computing the generalized binomial coefficients ${\kappa \choose [1,1]}$; however, the constant for the Jacobi polynomial complexity is considerably larger (our best guess sets it around $8$). 

The best asymptotical upper bound we can obtain using the inequalities \eqref{ine} is thus once again $O(k e^{\pi \sqrt{2k/3}})$.

The Jacobi parameters we chose for each of the computations below are $0$ and $1$.

\begin{table}[ht!]
\begin{center}
\begin{tabular}{|l||l||c||c||c|c|}
\hline
$k$ & $\kappa$ & Running time, & Running time, & Running time, & $U_{\kappa}$ \\ 
    &          & $\alpha = 1, ~m = l$ & $m$ symbolic & $\alpha, m$ symbolic & \\ \hline
$10$ & $[4,2,2,1,1]$ & $0.27$ & $0.74$ & $22.12$ & $42$ \\ 
     & $[4,3,3]$ & $0.11$ & $0.35$ & $1.88$ & $30$ \\
     & $[7,3]$ & $0.10$ & $0.30$ & $1.57$ & $26$ \\
\hline 
$15$ & $[6,4,2,2,1]$ & $1.05$ & $11.08$& $N/A$& $139$\\
     & $[3,3,3,3,3]$ & $0.39$ & $0.87$ & $63.07$ & $56$ \\
     & $[10,5]$ & $0.19$ &$1.01$  & $27.98$ & $51$\\
\hline
$20$ & $[6, 4, 3,2, 2,1,1,1]$ & $5.94$ & $N/A$ & $N/A$ & $418$\\
     & $[4,4,4,4,4]$ & $0.63$ & $8.24$  & $N/A$ & $126$ \\
     & $[12,8]$  & $0.26$ & $3.51$ & $N/A$ & $81$\\
\hline
$25$ & $[7,5,4,3,2,2,1,1]$ & $18.61$ & $N/A$ &$N/A$ & $1077$ \\
     & $[5,5,5,5,5]$ &$1.23$ &$N/A$ & $N/A$& $252$\\
     & $[16,9]$ & $0.45$ & $N/A$ & $N/A$ & $125$ \\
\hline
\end{tabular}
\end{center}
\caption{Running times (in seconds) for the Jacobi polynomial computation.} \label{table_jacob_times}\index{multivariate Jacobi polynomials!algorithm}
\end{table}

\begin{remark}
While the running times for numerical evaluation are reasonable, they explode when a symbolic parameter is introduced. The coefficients of the polynomial are rational functions of that parameter or combination of parameters, of order up to $k(k-1)/2$. We recall that there are $U_{\kappa, ~[1^2]}$ of them, a potentially superpolynomial number, which explains the tremendous increase in the running time.\end{remark}

%The algorithm has at its core recurrence (\ref{j}), and since the length of the recurrence is $k$, it follows that once again most of the work is done in computing binomial coefficients, and the bound is then the same, $O(k^6 P_k^2)$ or asymptotically $O(k^4 e^{2 \pi \sqrt{2/3} \sqrt{k}})$.

%The performance in practice is exemplified below.

%\begin{tabular}{|l|l|r|r|r|}
%\hline
%Partition sum & Partition &  \hspace{.25cm} Runtime    & \hspace{.25cm} Runtime & Ratio \\
%              &          & (in seconds) & (in seconds) & \\
%              &          & $\alpha = 2,g1 = 2, g1 = 1$ & symbolic $\alpha, g1, g2$ & \\ \hline
%$k=15$ & $\kappa = [14,1]$ & $2.82$ & $12.97$ & \\ 
%       & $\kappa = [8,7]$ & $0.95$ & $29.94$ & \\
%       & $\kappa = [3,3,3,3,3]$ & $1.73$ & $$ & \\ \hline
%$k=20$ & $\kappa = [19,1]$ &$7.75$ & $$ & \\ 
%       & $\kappa = [10,10]$ &$1.73$ & $$ & \\
%       & $\kappa = [5,5,5,5]$ &$5.44$ &$$ & \\ \hline
%$k=25$ & $\kappa = [24,1]$ &$16.42$ &$$ & \\ 
%       & $\kappa = [8,8,8,1]$ & $39.95$& $$ & \\
%       & $\kappa = [6,6,6,6,1]$ & $40.13$& $$ & \\ \hline
%$k=30$ & $\kappa = [29,1]$ & $33.10$& $$ & \\ 
%       & $\kappa = [10,10,10]$ & $16.56$& $$ & \\
%       & $\kappa = [6,6,6,6,6]$ & $41.13$& $$ & \\ \hline
%\end{tabular}

\item \textit{\textbf{laguerre}}.

We use the format given in Section \ref{format_laguerre}; it is easily established that the complexity of computing the Laguerre polynomial is dominated by the cost of computing the binomial coefficients, that is $O(klU_{\kappa, ~[1^2]})$, and once again the best asymptotical upper bound we can obtain using the inequalities \eqref{ine} is thus once again $O(k e^{\pi \sqrt{2k/3}})$.

The Laguerre parameter we chose for each of the computations below is $1$.

\begin{table}[ht!]
\begin{center}
\begin{tabular}{|l||l||c||c||c|c|}
\hline
$k$ & $\kappa$ & Running time, & Running time, & Running time, & $U_{\kappa}$ \\ 
    &          & $\alpha = 1, ~m = l$ & $m$ symbolic & $\alpha, m$ symbolic & \\ \hline
$10$ & $[4,2,2,1,1]$ & $0.12$ & $0.23$ & $0.54$ & $42$ \\ 
     & $[4,3,3]$ & $0.07$ & $0.14$ & $0.31$ & $30$ \\
     & $[7,3]$ & $0.07$ & $0.10$ & $0.28$ & $26$ \\
\hline 
$15$ & $[6,4,2,2,1]$ & $0.49$ & $0.82$  & $2.95$& $139$\\
     & $[3,3,3,3,3]$ & $0.18$ & $0.27$ & $0.84$ & $56$ \\
     & $[10,5]$ & $0.11$ &$0.22$  & $0.81$ & $51$\\
\hline
$20$ & $[6, 4, 3,2, 2,1,1,1]$ & $2.26$ & $3.37$ & $16.08$ & $418$\\
     & $[4,4,4,4,4]$ & $0.44$ & $0.69$  & $2.74$ & $126$ \\
     & $[12,8]$  & $0.20$ & $0.37$ & $1.79$ & $81$\\
\hline
$25$ & $[7,5,4,3,2,2,1,1]$ & $7.23$ & $11.06$ &$67.92$ & $1077$ \\
     & $[5,5,5,5,5]$ &$0.96$ &$1.53$ & $8.06$ & $252$\\
     & $[16,9]$ & $0.32$ & $0.69$ & $4.21$ & $125$ \\
\hline
\end{tabular}
\end{center}
\caption{Running times (in seconds) for the Laguerre polynomial computation.} \label{table_laguerre_times}
\end{table}
\index{multivariate Laguerre polynomials!algorithm}
\begin{remark} For the Laguerre polynomials, even in the all-symbolic case, the computation is very easy, and the storage required is relatively small. This explains why it is possible to obtain them without much effort, in any one of the cases.
\end{remark}

\item \textit{\textbf{hermite}}.

We use the format given in Section \ref{format_hermite} and recurrence \eqref{recurrence_hermite}. We only do work for those coefficients that correspond to subpartitions $\sigma$ of $\kappa$ such that $|\sigma| \equiv k~($mod$ 2)$. There are $A_{\kappa}$ of them. For each, we compute at most ${l \choose 2}$ contiguous coefficients, each computed with $O(k)$ complexity. The complexity of the rest of the computation is $O(k)$. Hence the total complexity is $O(kl^2A_{\kappa})$. 

\begin{remark} $A_{\kappa} = O(U_{\kappa})$; $A_{\kappa} \sim U_{\kappa}/2$.
\end{remark}

Hence one asymptotical upper bound that we can obtain for the complexity of computing a Hermite polynomial is $O(k^2 e^{\pi \sqrt{2k/3}})$.

\begin{table}[ht!]
\begin{center}
\begin{tabular}{|l||l||c||c||c|c|}
\hline
$k$ & $\kappa$ & Running time, & Running time, & Running time, & $A_{\kappa}$ \\ 
    &          & $\alpha = 1, ~m = l$ & $m$ symbolic & $\alpha, m$ symbolic & \\ \hline
$10$ & $[4,2,2,1,1]$ & $0.21$ & $0.24$ & $0.75$ & $22$ \\ 
     & $[4,3,3]$ & $0.09$ & $0.11$ & $0.33$ & $16$ \\
     & $[7,3]$ & $0.05$ & $0.06$ & $0.24$ & $14$ \\
\hline 
$15$ & $[6,4,2,2,1]$ & $0.41$ & $2.83$  & $42.92$& $88$\\
     & $[3,3,3,3,3]$ & $0.13$ & $0.17$ & $1.83$ & $38$ \\
     & $[10,5]$ & $0.10$ &$0.12$  & $1.10$ & $30$\\
\hline
$20$ & $[6, 4, 3,2, 2,1,1,1]$ & $1.93$ & $2.39$ & $N/A$ & $211$\\
     & $[4,4,4,4,4]$ & $0.35$ & $0.51$  & $N/A$ & $66$ \\
     & $[12,8]$  & $0.18$ & $0.25$ & $13.49$ & $43$\\
\hline
$25$ & $[7,5,4,3,2,2,1,1]$ & $6.23$ & $7.53$ & $N/A$ & $1077$ \\
     & $[5,5,5,5,5]$ &$0.90$ &$1.20$ & $N/A$ & $252$\\
     & $[16,9]$ & $0.29$ & $0.50$ & $106.56$ & $125$ \\
\hline
\end{tabular}
\end{center}
\caption{Running times (in seconds) for the Hermite polynomial computation.} \label{table_herm_times}\index{multivariate Hermite polynomials!algorithm}
\end{table}

\begin{remark} Note that when $m$ is parametrized, but $\alpha=1$, the computation is almost as fast as in the all-numerical case. That happens because the dependence on $m$ is very simple, and it only involves Pochhammer\index{Pochhammer symbol!see{ shifted factorial}} symbols, which do not get expanded (so that the storage required is minimal). However, the dependence on $\alpha$ is more complicated, and the rational functions obtained as coefficients are complex and hard to store. Hence the running time for the all-symbolic computation increases dramatically. \end{remark}

\end{enumerate}

\subsection{Conversion algorithms}

There are five conversion algorithms, \textbf{m2jack}, \textbf{jack2jack}, \textbf{m2m}, \textbf{p2m}, and \textbf{m2p}. 

\begin{enumerate}
\item \textbf{m2jack}. This algorithm computes and then inverts the change of basis matrix from monomials to Jack polynomials, taking advantage of the fact that the matrix is upper triangular. At each turn, the algorithm extracts the highest order monomial remaining, computes the coefficient of the corresponding Jack polynomial, and then extracts the monomial expansion of the Jack polynomial from the current monomial expression. 

Let $\kappa$ be the highest-order monomial present in the initial expression, and let us assume that the expression is homogeneous. 

Then the complexity of the computation is dominated by the complexity of computing the Jack polynomial expansion in terms of monomials for all partitions of $k$ smaller in lexicographical ordering than $\kappa$. 

It follows that an upper bound on the complexity is given by $O(D_{\kappa} k^4 D_{\kappa}) = O(k^2 e^{2 \pi \sqrt{2/3} \sqrt{k}})$. 

The performance in practice is exemplified below.

\begin{tabular}{|c|c|r|r|r|}
\hline
Partition sum & Partition                & Runtime & Runtime & Ratio \\
              &                          & $(\alpha=2)$ & symbolic $\alpha$ & of the two \\ \hline
$k=6$         & $\kappa = [6]$         & $0.14$ & $0.45$ & $0.31$\\ \hline
$k=7$         & $\kappa = [7]$           & $0.29$ & $1.04$ &  $0.27$ \\ \hline
$k=8$         & $\kappa = [8]$       & $0.63$ & $2.91$ & $0.21$ \\ \hline
$k=9$         & $\kappa = [9]$       & $1.21$ & $7.49$ & $0.16$ \\ \hline
$k=10$        & $\kappa = [10]$       & $2.62$ & $20.25$ & $0.12$ \\ \hline
$k=11$	      & $\kappa = [11]$	      & $4.77$  & $54.75$ & $0.08$ \\ \hline
$k=12$	      & $\kappa = [12]$	      & $8.82$  & $186.09$ & $0.04$  \\ \hline
$k=15$        & $\kappa = [15]$     & $52.65$ & $7177.02$ & $<0.01$ \\ \hline 
\end{tabular}

\item \textbf{m2m}. The algorithm takes an expression involving products of monomial functions and writes it in monomial basis by deciding which partitions appear in the expansion and by counting the number of times they appear. Hence this algorithm is an alternative to adding the $m$ basis as the dual of $h$ in the \textbf{SF} package, and using the \textbf{tom} procedure afterward (though the \textbf{tom} procedure is more general than this).

We have tested \textbf{m2m} against \textbf{tom}, and we have found that on partitions where the ratio sum-to-length is high, \textbf{m2m} performs much better, while on partitions with the ration sum-to-length is small, the tables are reversed. Hence we recommend to the user who wants to use our library, but might be working with partitions of the latter case, to also obtain and install \textbf{SF} and use it for computations. 

Below is a performance comparison. The number of variables $n$ used in \textbf{m2m}, each time, was the sum of the partition lengths (which is the smallest number of variables that requires obtaining all terms).

\begin{tabular}{|l|c|c|c|c|}
\hline
Input & $n$ & Runtime \textbf{m2m} & Runtime \textbf{tom} & Ratio \\ \hline
$m[5,2,1,1]\cdot m[4]$ & $5$ & $0.07$ & $0.28$ & $0.25$ \\ \hline
$m[3,2,1]\cdot m[5]$ & $4$ & $0.03$ & $0.10$ & $0.30$ \\ \hline
$m[5,3,2]\cdot m[4,3]\cdot m[2]$ & $6$  & $3.52$ & $35.27$ & $0.10$\\ \hline
$m[7,3,1]\cdot m[4,2]$ & $5$ & $0.30$ & $9.89$ & $0.03$ \\ \hline 
$m[4,3,2]\cdot m[6,4]$ & $5$ & $0.29$ & $35.72$ & $<0.01$ \\ \hline \hline
$m[2,2,1]\cdot m[1,1]$ & $5$& $0.05$ & $0.03$ & $1.66$ \\ \hline
$m[3,1]\cdot m[2,2,1]$ & $5$& $0.12$ & $0.05$ & $2.40$\\ \hline
$m[2,1,1,1]\cdot m[2,1]$ & $6$ & $0.22$ & $0.04$ & $5.50$  \\ \hline
$m[3,2,1]\cdot m[2,1,1,1]$ &$7$ & $2.95$ & $0.10$ & $29.5$ \\ \hline
$m[3,1,1]\cdot m[2,1]\cdot m[1,1,1]$ & $8$ & $13.55$ & $0.13$ & $104.23$ \\ \hline
\end{tabular}

\item \textbf{p2m}.
The algorithm expands a product of simple power sum functions into monomial basis. This is an alternative to adding the $m$ basis as the dual of $h$ in the \textbf{SF} package, and calling the \textbf{tom} procedure with power sum functions as inputs. As was the case with \textbf{m2m}, our algorithm performs much better on partitions with high sum-to-length ratio, and \textbf{tom} performs better on partitions with low sum-to-length ratio, as can be clearly seen from the performance comparison below.

\begin{tabular}{|l|c|c|c|}
\hline
Input & Runtime \textbf{p2m} & Runtime \textbf{tom} & Ratio \\ \hline
$p_4\cdot p_3^2\cdot p_2$ & $0.03$ & $0.14$ & $0.21$ \\ \hline
$p_8\cdot p_5\cdot p_1^3$ & $0.20$ & $5.46$ & $0.04$ \\ \hline
$p_7\cdot p_4\cdot p_3$ & $0.01$ & $0.46$ & $0.02$ \\ \hline
$p_5^2\cdot p_4^3\cdot p_2$ & $0.04$ & $5.40$ & $<0.01$ \\ \hline \hline
$p_4\cdot p_2^3\cdot p_1$ & $0.12$ & $0.10$ & $1.20$ \\ \hline
$p_5\cdot p_2^2\cdot p_1^3$ & $0.96$ & $0.17$ & $5.64$ \\ \hline
$p_3\cdot p_2\cdot p_1^5$ & $1.97$ & $0.04$ & $49.25$ \\ \hline
$p_2^4\cdot p_1^4$ & $16.27$ & $0.15$ & $108.46$ \\ \hline
\end{tabular}

\item \textbf{m2p}. The algorithm converts an expression of monomials into power sum functions; it is an alternative to the \textbf{top} option in the \textbf{SF} package. As before, for high sum-to-length ratio, our algorithm performs better, whereas the reverse is true for low sum-to-length ratio. It is perhaps worth noting that for this case, the ratio sum-to-length has to be between 1 and 2 for a significant outperformance of our \textbf{m2p} by \textbf{top} to occur. This can be seen in the performance examples below.

\begin{tabular}{|l|c|c|c|}
\hline
Input & Runtime \textbf{m2p} & Runtime \textbf{top} & Ratio \\ \hline
$m[1,1,1,1,1,1,1,1]$ & $3.61$ & $0.04$ & $90.25$\\ \hline
$m[3,2,2,2,1,1]\cdot m[2,1,1,1]$ & $2.19$ & $0.11$ & $19.91$ \\ \hline
$m[2,2,1,1,1]\cdot m[2,1]$ & $0.120$ & $0.03$ & $4.00$ \\ \hline
$m[1,1,1,1]\cdot m[1,1,1]$ & $0.03$ & $0.02$ & $1.50$ \\ \hline \hline
$m[4,3,2,2]$ & $0.06$ & $0.18$ & $0.33$ \\ \hline
$m[10,1]$ & $0.01$ & $0.10$ & $0.10$ \\ \hline
$m[5,4,3]\cdot m[3]^2$ & $0.02$ & $0.23$ & $0.08$ \\ \hline
$m[5,4,3,3]$ & $0.08$ & $3.21$ & $0.02$ \\ \hline
$m[3,3,2,1]\cdot m[5,2]$ & $0.07$ & $5.57$ & $0.01$ \\ \hline
\end{tabular}

\item \textbf{jack2jack}. This algorithm takes an expression in Jack polynomials, and turns it into a linear combination of Jack polynomials, by taking each multiplicative term, expanding it in monomial basis, then using \textbf{m2m} to get rid of the resulting multiplicative factors, and finally, \textbf{m2jack} to convert the linear combination of monomials back in Jack polynomial basis.

\begin{tabular}{|l|c|c|c|}
\hline
Input & $n$ & Runtime (symbolic) & Runtime ($\alpha = 1$) \\ \hline
$J[2,1]\cdot C[1,1,1]$& $3$ & $0.04$ & $0.03$ \\ \hline
$J[3,2,1]\cdot C[3]$ & $5$ & $3.53$& $0.45$ \\ \hline
$C[2]^2\cdot J[4,2]$ & $5$ & $11.44$ & $1.17$\\ \hline
$C[2]^2\cdot J[3,1]$ & $8$ & $29.09$ & $3.35$ \\ \hline
$P[3,2]\cdot J[2,1,1]$ & $7$ & $15.00$ & $2.10$\\ \hline
$P[2,1]\cdot J[4,2]\cdot C[2]$ & $5$ & $28.95$ & $2.07$ \\ \hline
\end{tabular}

\end{enumerate}

\subsection{Algorithms that evaluate integrals} \label{integr}

Here we have \textbf{expHjacks}, \textbf{expLjacks}, \textbf{expJjacks}, \textbf{expH}, \textbf{expL}, and \textbf{expJ}.

These algorithms depend on the length and complexity of the input. Let $P$ be the polynomial one wishes to analyze; one must first convert $P$ to a linear combination of Jack polynomials, and then replace each Jack polynomial with its expectation. 

\vspace{.25cm}

\noindent \textit{Case 1.} Suppose $P$ is in monomial format, as an expression which involves sums and products of monomials. First we convert $P$ to a linear combination of monomials using \textbf{m2m}, and then we convert that linear combination of monomials to a linear combination of Jack polynomials using \textbf{m2jack}. 

For any term of the form $m_{\lambda^1} m_{\lambda^2} \ldots m_{\lambda^p}$, with $\lambda^1, \lambda^2, \ldots, \lambda^p$ not necessarily distinct partitions, when we expand it in monomial basis, the largest possible number of terms is $D_{\mu}$, where $\mu$ is the partition which results from the superposition of $\lambda^1, \lambda^2, \ldots, \lambda^p$, i.e. $\mu_1 = \lambda^1_1 + \lambda^2_1+ \ldots + \lambda^p_1$, $\mu_2 = \lambda^1_2 + \lambda^2_2+ \ldots + \lambda^p_2$, etc.. Let $u = |\mu|$.

After the expansion in monomial basis, applying \textbf{m2jack} on the resulting expression has complexity $O(u^4 D_{\mu}^3) = O(u^2 e^{3 \pi \sqrt{2/3} \sqrt{u}})$. 

\begin{remark} This however is a very relaxed upper bound, and if we start off with $P$ being a sum of a few ($n$) monomials, the call to \textbf{m2m} is not executed, and the complexity of the call to \textbf{m2jack} is $O(n u^4 D_{\mu}^2) = O(n u^2 e^{3 \pi \sqrt{2/3} \sqrt{u}})$. 
\end{remark}

As explained in Section \ref{compexp}, the first step is common to \textbf{expH}, \textbf{expL}, and \textbf{expJ}. The second step is different and its complexity is much higher for \textbf{expH} than for \textbf{expL} or \textbf{expJ}. However, as we can see from the running times in the table below, the calls to \textbf{m2m} and \textbf{m2jack} (made in the first step) are much more expensive than the substitutions, and so the overall running times are comparable.

In these examples, we consider a symbolic parameter $a$, a symbolic number of variables $n$, $\gamma = 1$, and $g_1 = g_2 = 1$.

\begin{tabular}{|c|r|r|r|} 
\hline
Input & Runtime \textbf{expH} & Runtime \textbf{expL} & Runtime \textbf{expJ} \\ \hline
$m[6]$ & $0.94$ & $0.70$ & $0.80$ \\ \hline
$m[3,3,2]$ &$ 1.98$ & $0.85$  & $0.96$ \\ \hline
$m[5,2,1]$ &$5.69$ &$3.20$  & $4.20$ \\ \hline
$m[3, 1, 1, 1]\cdot m[2]$ & $4.23$ & $1.84$ & $2.59$ \\ \hline
$m[4,1]\cdot m[1,1,1]$ & $3.94$& $2.18$  &  $3.58$ \\ \hline
$m[5,1]\cdot m[2]$ & $8.86$ & $6.04$ & $9.82$ \\ \hline
$m[3]^2\cdot m[2]$ & $8.80$ & $7.00$ & $13.04$ \\ \hline
$m[4,2]\cdot m[3,1]$ & $39.85$ & $35.71$ & $68.82$ \\ \hline
\end{tabular}

\vspace{.25cm}

\noindent \textit{Case 2.} Suppose $P$ is in Jack polynomial format; then we use \textbf{jack2jack} to write the it as a linear combination of Jack polynomials, and finally we replace each Jack term by its expected value. The first step, as before, is common to all three procedures (\textbf{expHjacks}, \textbf{expLjacks}, \textbf{expJjacks}). 

While in the case of \textbf{expHjacks} the complexity of computing the expectation is $O(u^4 e^{2 \pi \sqrt{2/3} \sqrt{u}})$, in the cases of \textbf{expLjacks} and \textbf{expJjacks} the same complexity is only $O(u)$. This explains the significant differences recorded in the first three rows of the table. It is also worth noting that in the case of an odd $u$, the time it takes to compute the expected value of a Jack polynomial with Hermite weight is $0$, as the value of the output is known in advance to be $0$.

The complexity of expressing a product of Jack polynomials in Jack polynomial basis is much higher than the computation of a single Jack polynomial expected value. This explains why, in the last few rows of the table, the entries are no longer so different in magnitude.

In the examples below, we considered a symbolic parameter $a$, a symbolic number of variables $n$, $\gamma = 1$, and $g_1 = g_2 = 1$.

\begin{tabular}{|c|r|r|r|}
\hline
Input & Runtime \textbf{expHjacks} & Runtime \textbf{expLjacks} & Runtime \textbf{expJjacks} \\ \hline
$C[4,3,2,1]$ & $0.30$ &$0.03$  &$ 0.03$ \\ \hline
$C[6,6,2]$ & $1.06$ & $0.04$& $0.04$\\ \hline
$C[7,5,3,1]$ & $4.70$ &$0.04$  & $0.05$\\ \hline
$C[10, 3, 2, 1]$ & $4.47$ & $0.05$ & $0.05$ \\ \hline
$C[3,1]\cdot P[2,2]$ & $14.75$ & $12.75$ & $12.93$ \\ \hline 
$C[4,2]\cdot J[1,1]$ & $31.86$ & $29.05$ & $30.11$ \\ \hline
$J[2,1,1]\cdot J[4]$ & $76.62$ & $81.93$ & $80.14$ \\ \hline
$C[2,2]\cdot J[4]$ & $53.79$ & $54.30$ & $55.07$ \\ \hline
\end{tabular}

\vspace{.25cm}

\subsection{Numerical algorithms}

Some of the symbolic/numerical evaluation routines analyzed in the previous sections include options for polynomial evaluation on numerical values of the $x$ variables. The routines that compute the polynomials Jack, Hermite, Laguerre, and Jacobi have options that allow for numerical values of the $x$ variables. This makes it possible to compute quantities like $C_{[3,2]}^{3}(2.53, -1.09, 7.33)$; this feature can be used for graphics (when one needs to plot some statistic of a random matrix, as we demonstrate in the next section).

The algorithms we have used to implement these options have been developed and analyzed by Koev and Demmel \cite{plamen_jacks} for the Jack polynomials; to evaluate the other polynomials, we use the regular expansion in terms of Jack polynomials, then substitute the numerical values for each Jack polynomial.

\section{Applications}

We have written this library for the user who would like to do statistical computations, form or test conjectures, and explore identities. The great benefit is that all computations can be done symbolically, keeping $\alpha$ as a parameter; the downside of symbolic computations, as we have mentioned before, is that the storage space required is very large, and computations are consequently slowed down. Our experience, however, was that on a good, but not top-of-the-line machine (see specifications in Section \ref{perf}), we have been able to increase the size of the partition enough in order to make and then satisfactorily test conjectures.

Below are some examples of computations that we imagine are of the type a researcher might want to use in forming conjectures, or of the type that might be useful in practice.

Some of the applications, like the computation of the moments of the trace, can be done with symbolic $\alpha$ and $n$ (number of variables); others, like the computation of the moments of the determinant, need an actual value for $n$, but allow for symbolic $\alpha$ computations; yet others, like the level density computation, need all numerical parameters. For each computation, we have tried to indicate upper bounds for the size of the necessary numerical parameters.

\begin{enumerate}
\item \textit{Moments of the determinant.} One of the many interesting problems in random matrix theory is computing the moments of the determinant of a square random matrix. If the eigenvalues are chosen to have the $2/\alpha$-Hermite distribution (given by the weight function $\mu_{H}^{\alpha}$), the problem of computing the determinant is non-trivial. Closed form answers are known for the cases $\alpha = 1/2, 1$, and $2$ (see  \cite{jackson_det},  \cite{Del-Lec}, \cite{mehta_moments}); however, the general $\alpha$ case does not have an explicit answer (except for some particular situations like in \cite[chapter 8]{dumitriu03th}. 

Since the $k$th moment of the determinant's distribution is given as the integral of $m_{[k^m]}(x_1, \ldots, x_m) = C_{[k^m]}^{\alpha}(x_1, \ldots, x_m)/ C_{[k^m]}(I_m)$ over the corresponding $2/\alpha$-Hermite distribution, \textbf{MOPS} can be used in evaluating it for specific values of $k$ and $m$.

For example, for $k = 2$ and $m = 5$, the answer can be obtained by typing in

\footnotesize{{\bf 
$~~~~~~$ {\mathversion{bold}$>$} factor(expHjacks(a, C{\mathversion{bold}$[$}2,2,2,2,2{\mathversion{bold}$]$}, 5){\mathversion{bold}$/$}jackidentity(a, {\mathversion{bold}$[$}2,2,2,2,2{\mathversion{bold}$]$}, 5));}}

\normalsize
and the output is
\footnotesize{{\mathversion{bold}
\[
>~~~~~~~~~~~~~~~~~~~\frac {\mbox{{\bf a}}^{4}+10\,\mbox{{\bf a}}^{3}+45\, \mbox{{\bf a}}^{2}+80\,\mbox{{\bf a}}+89}{\mbox{{\bf a}}^{4}}~~~~~~~~~~~~~~~~~~~~~~~~~~~~~~~~~~~~~~~~~~~
\] }}
\normalsize
The duality principle between $\alpha$ and $1/\alpha$ proved in \cite[Section 8.5.2]{dumitriu03th} linking the expected value of the $k$th power of the determinant of a $n \times n$ matrix to the expected value of the $n$th power of a $k \times k$ matrix is illustrated below:

\footnotesize{{\bf 
$~~~~~~$ {\mathversion{bold}$>$} factor(expHjacks({\mathversion{bold}$1/$}a, C{\mathversion{bold}$[$}5,5{\mathversion{bold}$]$}, 2){\mathversion{bold}$/$}jackidentity({\mathversion{bold}$1/$}a, {\mathversion{bold}$[$}5,5{\mathversion{bold}$]$}, 2));}}

\normalsize with output

\footnotesize{{\mathversion{bold}
$~~~~~~~ > ~~~~~~~~~~~~~~~ -\mbox{{\bf a}} \left( \mbox{{\bf a}}^{4}+10\,\mbox{{\bf a}}^{3}+45\, \mbox{{\bf a}}^{2}+80\,\mbox{{\bf a}}+89 \right)$}}

%
%Similarly, for any $\alpha$, $k=8$, $m=3$, the answer is
%\[
%>{\frac { 9~\left( 9\,{\alpha}^{6}+63\,{\alpha}^{5}+230\,{\alpha}^{4}+271\,{\alpha}^{3}+
%230\,{\alpha}^{2}+63\,\alpha+9 \right)  \left( {\alpha}^{2}+\alpha+1 \right) }{{\alpha}^{8}}}
%\]
%while for $1/\alpha$, $k=3$, m = $8$, we obtain
%\[
%\]
\normalsize

\begin{remark} In practice, we have observed that computations with $\alpha$ symbolic and $k \cdot m \leq 22$ can be performed relatively fast (under 2 minutes on the computer with specs given in the beginning of Section \ref{perf}); for $k \cdot m > 22$ and $\alpha$ symbolic, the amount of memory available begins to play an important role. For actual values of $\alpha$ (for example, $\alpha= 1$), the computation for $k=10$ and $m = 5$ took under 40 seconds. \end{remark}

\item \textit{Expectations of powers of the trace.}
Consider the problem of computing the expected value of the $6$th power the trace of a Hermite (Gaussian) ensemble (here $n$ is an arbitrary integer). This amounts to making a call to \textbf{expH}, simplifying, and expanding the answer in Taylor series for a clear format. In short, a one-line command:

\footnotesize{{\bf 
$~~~~~~${\mathversion{bold} $>$} taylor(simplify(expH(a, 
m{\mathversion{bold} $[$}6{\mathversion{bold}$]$, n})), n); }}

\normalsize
with answer
\footnotesize{
{\mathversion{bold} 
\[
~~~~~~>~~\frac{15\,\mbox{{\bf a}}^{3}-32\,\mbox{{\bf a}}^{2}+32\, \mbox{{\bf a}}-15}{\mbox{{\bf a}}^{3}} \mbox{{\bf n}}+ \frac{-54\, \mbox{{\bf a}}+ 32\,\mbox{{\bf a}}^{2}+32}{\mbox{{\bf a}}^{3}} \mbox{{\bf n}}^{2}+\frac {22\, \mbox{{\bf a}}-22}{\mbox{{\bf a}}^{3}} \mbox{{\bf n}}^{3}+\frac{5}{\mbox{{\bf a}}^3} \mbox{{\bf n}}^{4}~.
\]}}
\normalsize

\begin{remark} This computation emphasizes best the power of MOPs. It is very quick (took $0.8$ seconds on the test machine (see specifications in Section \ref{perf}) and it allows for both $\alpha$ and $n$ symbolic. The same computation for the 12th power of the trace with $\alpha$ and $n$ symbolic took less than $8$ minutes. \end{remark}

Integrals of powers of the trace are related to Catalan numbers and maps on surfaces of various genuses, and are of interest to (algebraic) combinatorialists (\cite{Forrester_book, goulden_jackson}).

\item \textit{Smallest eigenvalue distributions.} 
One of the quantities of interest in the study of Wishart matrices\footnote{The joint eigenvalue distribution of Wishart matrices is given by the Laguerre weight $\mu_{L}^{\alpha, \gamma}$ with $\alpha = 1$ (complex case) or $\alpha = 2$ (real case).} is the distribution of the smallest eigenvalue. There is an extensive literature on the subject, starting with the work of James \cite{james64a} and Constantine \cite{constantine_noncentral}. More recent references are Silverstein \cite{silverstein89a} and Edelman \cite{edelman91a}. In \cite{dumitriu03th}, we find a closed-form answer for general $\alpha$ and integer values of $\gamma$, in terms of a hypergeometric ${}_2F_{0}$ function (see also \eqref{seld}). 

We wrote a small script (presented below) implementing the formula, and used it
to compute the exact distribution of the smallest eigenvalue of a Wishart matrix
for $\alpha= 1$ (the complex case) for $n = 3, ~m =6$, and $n = 2,~m=10$, which
we plotted in MATLAB. We have also used a Monte Carlo simulation to plot in
MATLAB histograms of the smallest eigenvalue of matrices from the 
corresponding
Wishart ensemble, for comparison (see Figure \ref{smaleig}). For the 
histograms, 
we have chosen in each case $30,000$ samples from the corresponding Wishart 
ensemble.

\vspace{.25cm}

\footnotesize{
{\bf smalleig:=proc(n,k,x) local r,t,i, y,inte; \\
$~~~~~$ if (n$>$1) then r:=[-2/x]; \\ 
$~~~~~$ end if; \\
$~~~~~$ for i from 2 to (n-1) do \\
$~~~~~~~~~~$ r:=[op(r),-2/x]; \\
$~~~~~$ end do; \\
$~~~~~$ t:=x\^{}((k-n)*n) * exp(-x*n/2) * ghypergeom(1, [n-k, n+1],[],r,'m'); \\
$~~~~~$ return simplify(t); \\
end proc; \\

\vspace{.25cm}

scaledsmalleig:=proc(n,k,x) local inte, yy, z; \\
$~~~~~$ yy :=z-$>$smalleig(n,k,z); \\
$~~~~~$ inte := integrate(yy(z), z=0..infinity); \\
$~~~~~$ return(smalleig(n,k,x)/inte); \\
end proc; \\

\vspace{.25cm}

zz:=scaledsmalleig(3,6, x); \\
plot(zz, x=0..10); \\

}}

\normalsize

\begin{figure}[ht]
\parbox[b]{5cm}{\epsfig{figure= 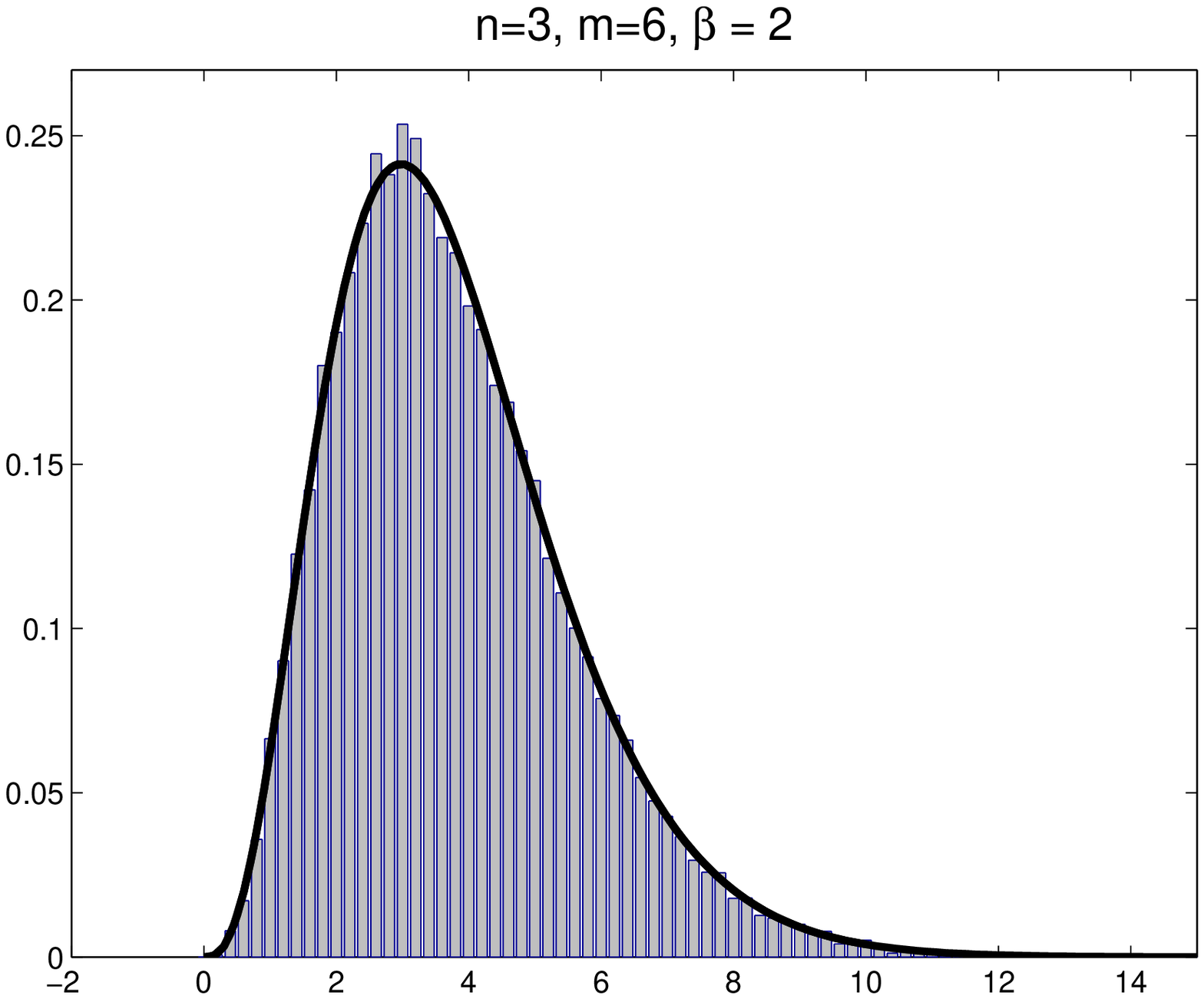, height = 5cm}} \hspace{2cm}
\parbox[b]{5cm}{\epsfig{figure=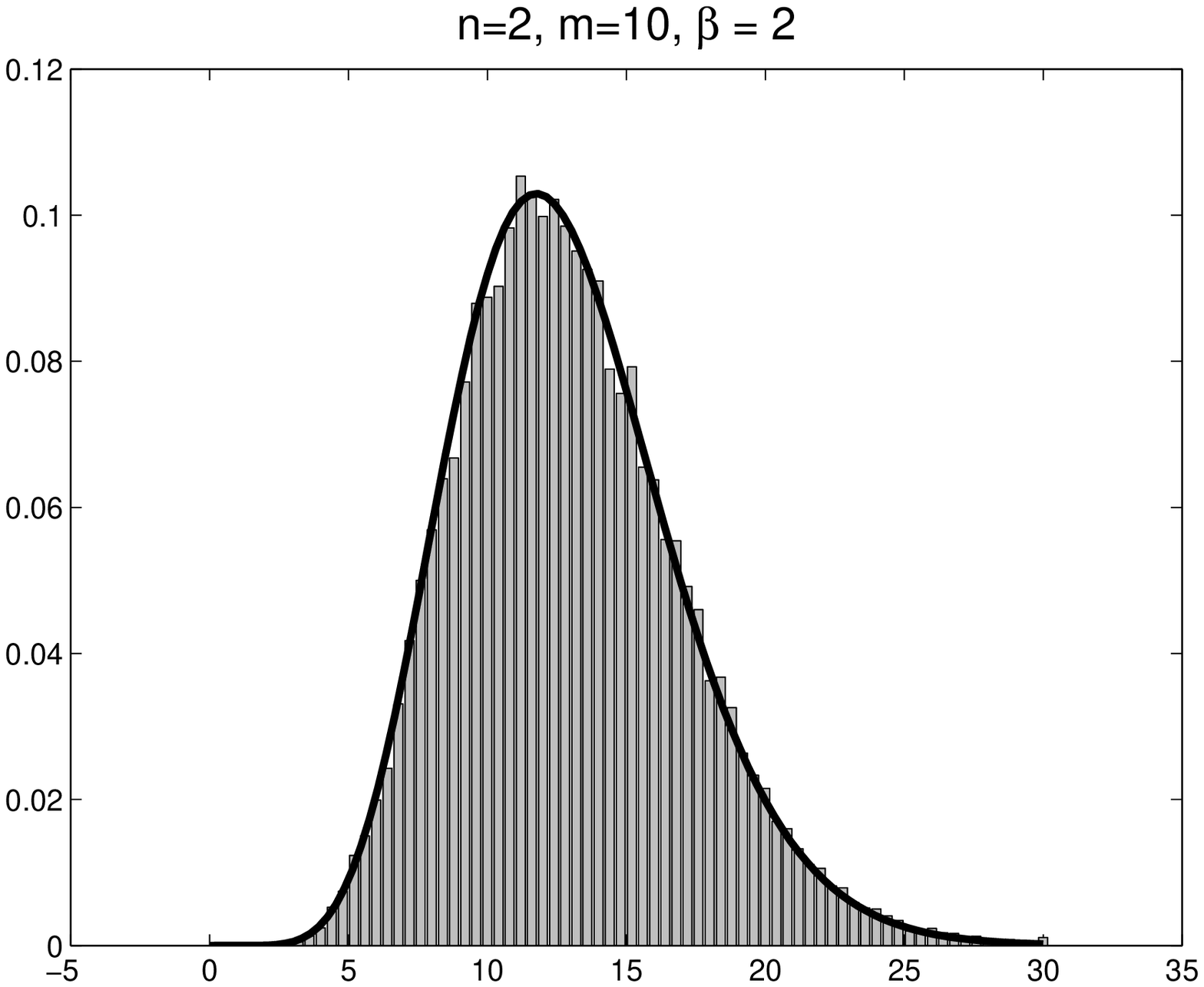, height = 5cm}} 
\caption{Histograms of the smallest eigenvalue distribution for the complex Wishart 
ensembles of size $(3,6)$ and $(2, 10)$ ($\alpha = 1$), together with the exact 
distributions as given by \eqref{seld}.} \label{smaleig} \end{figure}

\item \textit{Level densities.} Level density formulas are well-known in
terms of orthogonal polynomials for $\alpha = 1/2, 1, 2$. Forrester and
Baker \cite{Forrester_poly} have computed these densities in terms of a
multivariate Hermite polynomial for $\beta = 2/\alpha$ an even integer
(i.e. $\alpha$ is the inverse of an integer). We have found an equivalent
formulation for the level density of the $n \times n$ Hermite ensemble for which $\alpha$ is the inverse of an integer (equivalently, $\beta = 2/\alpha$ is an even integer). This formula is presented below:
\[
\rho_n(x) ~=~ \frac{1}{\sqrt{2 \pi}}(-1)^{n/\alpha} \frac{\Gamma \left(1+\frac{1}{\alpha} \right)}{\Gamma \left( 1+ \frac{n}{\alpha} \right)} e^{-x^2/2} ~H_{[(2/\alpha)^{n-1}]}^{\alpha} (xI_{n})~,
\]
where the partition $[(2/\alpha)^{n-1}]$ is the partition that consists of $2/\alpha$ repeated $n-1$ times.

To compute the Hermite polynomial, we used the formula \eqref{mine}, and in order to get all the eigenvalues roughly in $[-1, 1]$ we scale both the variable and the density function by $\sqrt{2n\beta} \equiv \sqrt{4n/\alpha}$ (see the script).

We have used the script below
to produce Figure \ref{densities}, which is an exact plot of the level 
densities
for $n = 4$, and $\beta = 2,4,6,8,10$ (equivalently, $\alpha = 1, 1/2,
1/3, 1/4, 1/5$).

\vspace{.5cm}

\footnotesize{
{\bf
leveldens:=proc(a,k::list, n, x) option remember; \\
$~~$ local s,u,ut,ul,ks,ss,j,i,sp,result,t,t1,r,jp,ul1,c,bbb; \\
$~~$ if(not(`MOPS/parvalid`(k))) then return; \\
$~~$ end if; \\
$~~$ result:=0; ks:=sum(k[i],i=1..nops(k)); sp:=`MOPS/subPar`(k); 

\#\# we compute the Hermite polynomial evaluated at $xI_n$, using formula 
\eqref{mine} 

$~~$ for s in sp do \\
$~~~~~$ ss:=0; c:=0; ss:=sum(s[i],i=1..nops(s)); \\
$~~~~~$ if not((ss mod 2) = (ks mod 2)) then next; \\ 
$~~~~~$ end if; \\
$~~~~~$ for j from ss to (ks+ss)/2 do \\
$~~~~~~~$ jp:=`MOPS/Par`(j); \\
$~~~~~~~$ ul1:=(convert(jp,set) intersect convert(sp,set)); \\
$~~~~~~~$ ul:=[]; \\
$~~~~~~~$ for ut in ul1 do	\\
$~~~~~~~~~~$ if `MOPS/subPar? \hspace{-.15cm}`(s,ut) then ul:=[op(ul),ut];\\
$~~~~~~~~~~$ end if; \\
$~~~~~~~$ end do; \\
$~~~~~~~$ t:=0; \\
$~~~~~~~$ for u in ul do \\	
$~~~~~~~~~~~$ 
t1:=`MOPS/GSFact`(a,r+(n+a-1)/a,k)/`MOPS/GSFact`(a,r+(n+a-1)/a,u); \\
$~~~~~~~~~~$ t:=t+`MOPS/GBC`(a,k,u)*`MOPS/GBC`(a,u,s)*coeff(t1,r,(ks+ss)/2-j); \\
$~~~~~~~$ end do; \\
$~~~~~~~$ c:=c+t*(-1)\^{}j; \\
$~~~~~$ end do; \\
$~~~~~$ bbb:=factor(c*(-1)\^{}(ss/2)*x\^{}(ss));  \\
$~~~~~$ result:=result+bbb; \\
$~~$ end do; \\
$~~$ result:= result*(-1)\^{}(ks)*(-1)\^{}(ks/2) * exp(-x\^{}2/2) * 
1/sqrt(2*Pi); \\
$~~$ result:=result * factor(GAMMA(1+1/a)/GAMMA(1+m/a)); \\
end proc;

\vspace{.25cm}

\#\# we scale both the variable and the density function by $\sqrt{2n\beta}$

z:=(x,b)-$>$sqrt(2*4*b)*leveldens(2/b, [b,b,b], 4, x*sqrt(2*4*b));

\vspace{.5cm}

plot({z(x,2), z(x,4), z(x,6), z(x,8), z(x,10)}, x=-1.2..1.2, y=-.1..1.4);
}}

\vspace{.5cm}

\normalsize
\begin{figure}[ht!]
\caption{Level densities for $n=4$, $\alpha = 1, 1/2, 1/3, 1/4, 1/5$; 
``bumps'' increase as $\alpha$ decreases.} \label{densities}
\begin{center}
\psfig{figure=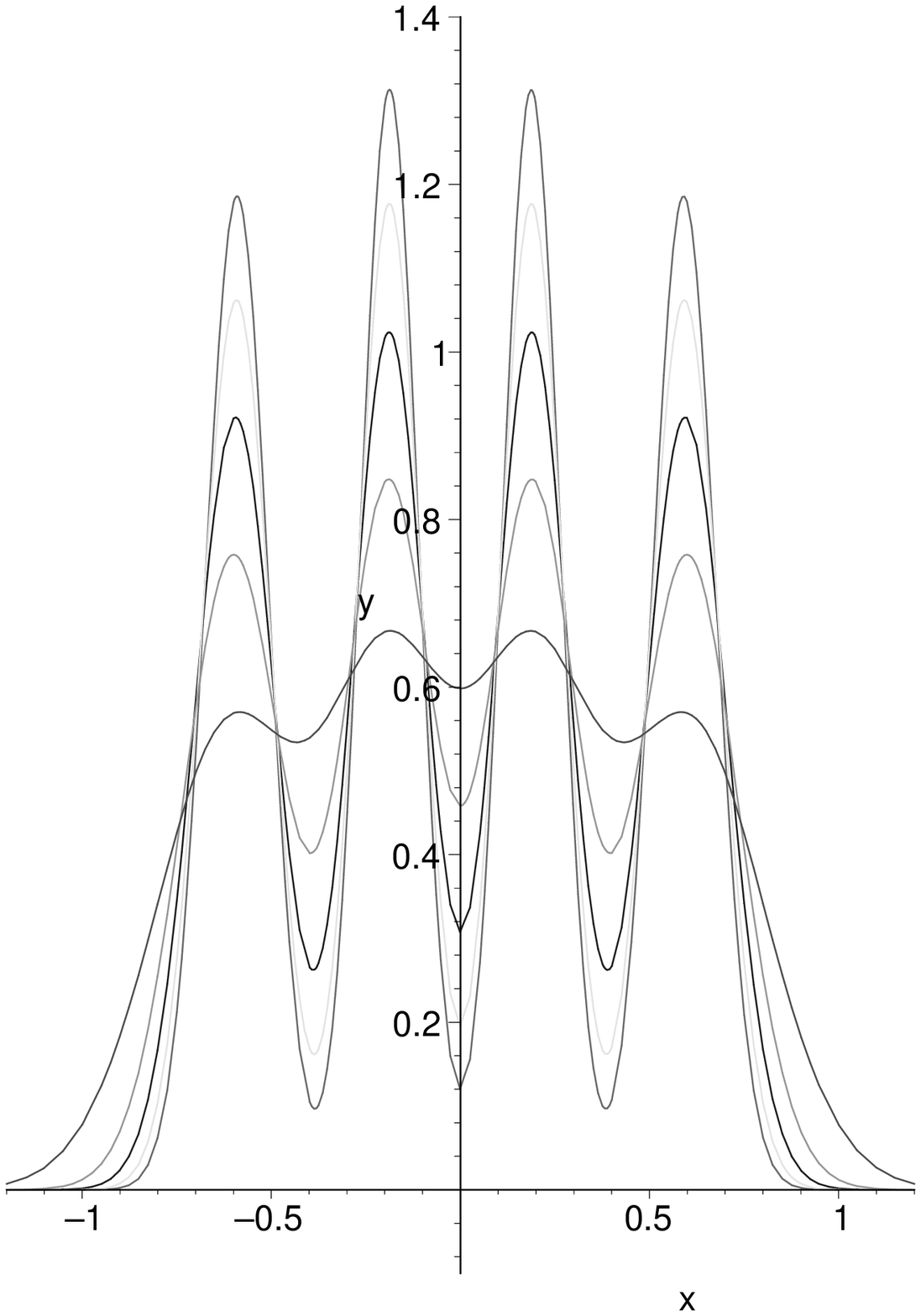, height = 7.5cm, width = 8cm}
\end{center}
\end{figure}

\vspace{3cm}

For illustration purposes, here is the exact (scaled) density for $\alpha = 1/4$ and $n = 5$, plotted above:
\footnotesize{
{\mathversion{bold} \begin{displaymath}
{\renewcommand{\arraystretch}{1.3}
  \begin{array}{l}
    \multicolumn{1}{c}{{\dfrac{\sqrt{10}{e^{-40\,{x}^{2 }}}}
        {50685458503680000\sqrt{\pi}}}\, \times}\\[1.7ex] 
    \hbox to 0pt{\hss$($}
    2814749767106560000000000000000\,{x}^{32}-
    2814749767106560000000000000000\,{x}^{30}+\\
    1720515795143884800000000000000\,{x}^{28}-
    696386684568207360000000000000\,{x}^{26}+\\
    194340604354756608000000000000\,{x}^{24}-
    36625240845346406400000000000\,{x}^{22}+\\
    4740055701777285120000000000\,{x}^{20}-
    658121972672102400000000000\,{x}^{18}+\\
    162266873453346816000000000\,{x}^{16}-
    31084533121233715200000000\,{x}^{14}+\\
    2673909486122434560000000\,{x}^{12}-
    136819200341311488000000\,{x}^{10}+\\
    29341248756019200000000\,{x}^{8}-
    1130060455927603200000\,{x}^{6}+\\
    67489799891754240000\,{x}^{4}-
    2060099901411552000\,{x}^{2}+32632929952848225).
  \end{array}}
\end{displaymath}}}

\normalsize
\item \textit{Conjectures.} We present here a conjecture that we formulated with the help of MOPs. This conjecture was proved later by Richard Stanley.

%We show here a screenshot of the library, which has lead to the following conjecture, later tested more thoroughly.

%\begin{figure}[ht] \label{screenshot}
%\begin{center}
%\epsfig{figure=conj.eps, height = 10.5cm}
%\caption{Illustrating a conjecture}
%\end{center}
%\end{figure}

\begin{Conjecture} Let $k$ be an integer, $\alpha$ a positive real, and consider the representation of the monomial function 
\[
m_{[k]} = \sum_{\lambda \vdash k} f_{\lambda, \alpha} C_{\lambda}^{\alpha}~.
\]
Then for all $\lambda$
\[
f_{\lambda, \alpha} = \frac{1}{n(\lambda)} \prod_{i=1}^{length(\lambda)} 
\left ( - \frac{i-1}{\alpha} \right)_{\lambda_{i}}~,
\]
where $n(\lambda)$ is an integer which does not depend on $\alpha$.
\end{Conjecture}

\end{enumerate}

\section{Copyleft}

Copyleft 2004 Ioana Dumitriu, Alan Edelman, and Gene Shuman.

Permission is granted to anyone to use, modify, and redistribute MOPs freely, subject to the following:
\begin{itemize} \item We make no guarantees that the software is free of defects.
\item We accept no responsibilities for the consequences of using this software.
\item All explicit use of this library must be explicitly represented. 
\item No form of this software may be included or redistributed in a library to be sold for profit without our consent.
\end{itemize}

\bibliography{bib_10_01}
\bibliographystyle{plain}

\end{document}